\documentclass[apjl]{emulateapj}
\usepackage{graphicx}

\slugcomment{To appear in The Astrophysical Journal}

\shorttitle{COMPLETE Outflows in Perseus}
\shortauthors{Arce, et al.}

\begin{document}


\title{The {\it COMPLETE} Survey of Outflows in Perseus}


\author{H\'{e}ctor G. Arce}
\affil{Department of Astronomy, Yale University, P.O. Box 208101, New Haven CT 06520}
\email{hector.arce@yale.edu}

\author{Michelle A. Borkin}
\affil{School of Engineering and Applied Sciences, Harvard University, 29 Oxford Street, Cambridge MA 02138}

\author{Alyssa A. Goodman}
\affil{Harvard-Smithsonian Center for Astrophysics, 60 Garden Street, Cambridge MA 02138}

\author{Jaime E. Pineda}
\affil{Harvard-Smithsonian Center for Astrophysics, 60 Garden Street, Cambridge MA 02138}

\and

\author{Michael W. Halle}
\affil{Surgical Planning Lab, Department of Radiology, Brigham and Women's Hospital, 75 Francis Street, Boston MA 02115}
\affil{Initiative in Innovative Computing, Harvard University, 60 Oxford Street, Cambridge MA 02138}


\begin{abstract}

We present a study on the impact of molecular outflows in the Perseus molecular cloud complex using the COMPLETE survey large-scale $^{12}$CO(1-0) and $^{13}$CO(1-0) maps. We used three-dimensional isosurface models generated in RA-DEC-Velocity space to visualize the maps. This rendering of the molecular line data allowed for a rapid and efficient way to search for molecular outflows over a large ($\sim 16$ deg$^2$) area. Our outflow-searching technique detected previously known molecular outflows as well as new candidate outflows. Most of these new outflow-related high-velocity features lie in regions that have been poorly studied before. These new outflow candidates more than double the amount of outflow mass, momentum, and kinetic energy  in the Perseus cloud complex. Our results indicate that outflows have significant impact on the environment immediately surrounding localized regions of active star formation, but lack the energy needed to feed the observed turbulence in the {\it entire} Perseus complex. This implies that other energy sources, in addition to protostellar outflows, are responsible for turbulence on a global cloud scale in Perseus. We studied the impact of outflows in six regions with active star formation within Perseus of sizes in the range of 1 to 4 pc. We find that outflows have enough power to maintain the turbulence in these regions and enough momentum to disperse and unbind some mass from them. We found no correlation between outflow strength and star formation efficiency for the six different regions we studied, contrary to results of recent numerical simulations. The low fraction of gas that potentially could be ejected due to outflows suggests that additional mechanisms other than cloud dispersal by outflows are needed to explain low star formation efficiencies in clusters.

\end{abstract}




\section{Introduction}
\label{introsection}

Outflows are an intrinsic part of the star formation process, as all stars, both low and high mass, go through a mass-loss phase during their protostellar stages (e.g., Arce et al.~2007).  
The outflowing supersonic wind from a protostar can accelerate the surrounding molecular gas to velocities significantly greater than those of the quiescent cloud gas thereby producing a molecular outflow.  
Early on in the study of outflows from young stars it was realized that they have the potential to have a major effect on the dynamics and structure of their parent cloud \citep{norman80}.  More recently, it has been suggested that in regions
of low-mass star formation, outflows
could be the leading disruptive agent that limits the lifetime of their parent molecular cloud \citep{hartmann01}.  
Analytical and numerical studies indicate that outflows can couple
strongly to the cloud and are highly efficient at driving turbulent motions 
(Matzner 2007; Nakamura \& Li 2007; Cunningham et al. 2009; Carroll et al.~2009)
and can also regulate the cloud's star formation efficiency (e.g.,  Matzner \& McKee 2000;  Nakamura \& Li 2007). 
Yet, other numerical studies suggest
that protostellar outflows do a poor job at driving cloud turbulence, but can disrupt dense clumps and affect the
cloud structure (Banerjee et al. 2007). 
These studies show the increased attention that research on the impact of outflows
has obtained, as well as the need for targeted observations required to constrain the models and to reconcile their discrepancies.

 The underlying physics in studies of outflow-environment interactions is mostly understood (even if difficult to model), however observations are crucial for constraining the various assumptions made in these models (e.g., outflow power, degree of collimation, mass outflow rate, outflow lifetime, etc.).  Significant constraints on the values of outflow characteristics require systematic observations of large outflow samples. Different physical processes triggered by protostellar outflows may be traced at different wavelengths (see, e.g., Hartigan et al.~2000). Shock-excited infrared lines of H$_2$ and optical Herbig-Haro (HH) objects trace the recently shocked gas that cools within a few years. On the other hand,  
 the high velocity CO  trace the surrounding molecular gas entrained by the protostellar wind and remains visible by collisional excitation for much more time than the shocked gas. This is why CO has mostly been used for studying the impact of outflows on the surrounding molecular cloud.  
  However, the CO  fails as an outflow tracer  in regions where there is little molecular gas (i.e., in the outskirts of the cloud), and so molecular outflows give a lower limit on the outflow momentum and energy injection into their (low-density) surroundings.   In these low-density regions other tracers like the HI (21~cm) and  CII (157 $\micron$) lines should be useful for tracing the outflow.


Observational studies have shown that outflows, even from low-mass stars, can have an impact on their cloud at different distances from the source ranging from a few thousand AU to several parsecs.  Survey studies of the circumstellar gas within $10^4$ AU of low-mass protostars indicate outflows contribute significantly to the mass-loss of the surrounding dense gas \citep{fuller02,arce06}.  An outflow's impact on its parent core, at distances of about 0.1 to 0.3~pc from the forming star, is evidenced through the detection of outflow-blown cavities and fast-moving (outflowing) dense gas  \citep[e.g.,][]{tafalla07}.  Giant outflows from young stars with sizes exceding 1~pc in length are common \citep{reipurth97,stanke00}.  These outflows can interact with the surrounding medium and induce changes in the velocity and density distribution of the parent cloud's gas at large distances from the source \citep{arce01b,arce02a}.  Millimeter studies show that many molecular outflows produced by a group or cluster of young stars interact with a substantial volume of the cluster's environment, may sweep up the gas and dust into ``shells'' and can have the energy required for driving the turbulence in the cluster gas \citep[e.g.,][]{knee00,williams03,stanke07}.  All these observational studies suggest that individual, groups, and clusters of outflows have a significant impact on their surroundings within a few parsecs from their sources.  The collective impact of all the outflows on an entire molecular cloud or on a molecular cloud complex (with size $\sim $10~pc) in which they reside is, however, still unclear.

\begin{figure*}
\epsscale{1.3}
\plotone{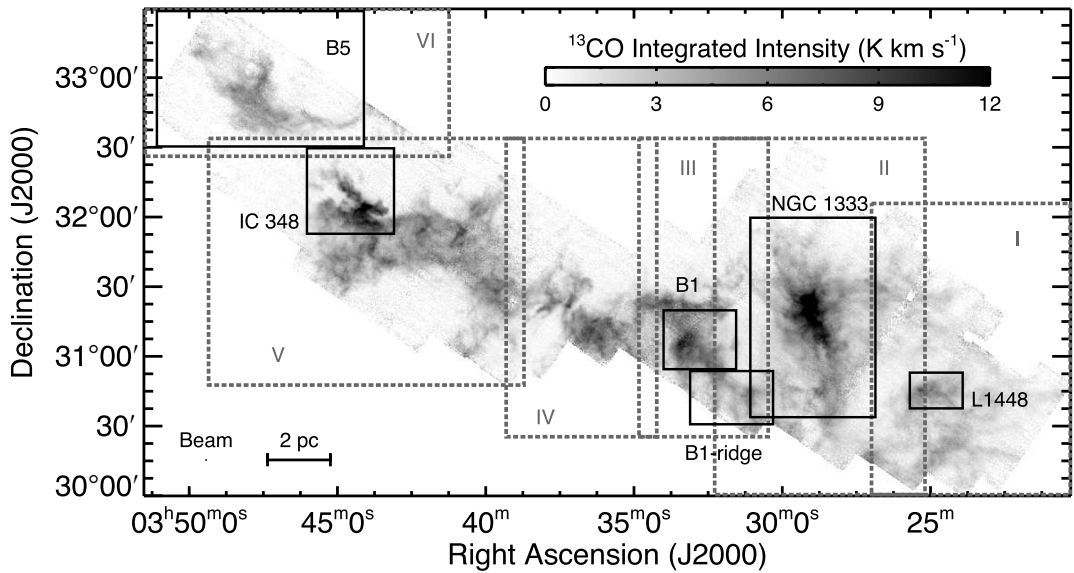}
\vspace{-3in}
\caption{$^{13}$CO(1-0) integrated intensity map of the  Perseus molecular cloud complex from the COMPLETE survey (Ridge et al.~2006a).
 The six areas visualized in 3D Slicer are labeled I through VI. The black (solid line) boxes show the approximate boundaries of the star-forming regions used in our analysis in section~\ref{analysis}  
 (see also Tables~\ref{regiontab} and \ref{physregtab}).
 Note that the $^{13}$CO map is not corrected for the FCRAO beam efficiency.
 \label{perseus_map}}
\end{figure*}

Large-scale (unbiased) molecular outflow surveys in the 1980's were crucial in estimating the molecular outflow energy in whole clouds, and acknowledging that outflows can have a significant impact on their host cloud.  These surveys were conducted toward clouds with high-mass star formation ---i.e., Mon OB1 \citep{margulis86} and the Orion southern cloud \citep{fukui86}--- using small (4 to 5~m) millimeter telescopes resulting in large beams (2.3\arcmin \/ to 2.7\arcmin).  Only regions with detectable high-velocity gas ($v > 10$~km~s$^{-1}$) were then mapped at higher (sub-arcminute) resolution.  Beam dilution should have hampered the ability to detect small ( $< 0.4$~pc) and slow ($v < 10$~km~s$^{-1}$) outflows in these surveys and many such outflows were probably missed.  Comparing recent higher angular resolution observations of small regions included within the large-scale surveys conducted about two decades ago, it is clear that these surveys detected only the most powerful outflows and that individual outflows were not resolved in high-density regions; 
 compare, for example, the results of \citet{yu00} and \citet{fukui86}.

Most of the recent outflow studies are restricted to small fields, focusing on individual objects or small dense regions of known star formation activity, thereby providing a limited view of the outflows' impact on the environment at large (cloud) scales.  These biases inhibit a full census of the outflows in a cloud, prevent the identification of multi-parsec outflows, and limit the study of the outflows' impact on the entire  molecular cloud complex.  For example, the studies of \citet{hat07} and \citet{hat09} present a search for outflows towards known cores in the Perseus molecular cloud complex, and the study succeeds in deriving outflow properties of outflows in the Perseus cloud in a consistent way and finding a few new (previously undetected) outflows.  However, they fail to provide an unbiased survey of outflows across the entire Perseus cloud complex as most observations are concentrated within  $\sim0.15$~pc from known cores and regions of active star formation. In contrast, the outflow survey presented here covers the full extent of the Perseus molecular cloud complex, and has the advantage of being able to provide information on how individual and groups of outflows effect the dynamics of the gas in the entire cloud complex and how they interact with their surroundings at different distances from the driving source (from about 0.06~pc to a few parsecs).


The Perseus molecular cloud complex is a chain of clouds with a total mass of a few thousand solar masses, and encompasses a total area of about 70~pc$^{2}$ \citep{evans09}. There is a span of distance estimates to Perseus that range from 230~pc \citep{cer90} to 350~pc \citep{her83}, most probably because different regions of the cloud complex are at different distances along the line of sight, and thus 
using a single distance for the entire cloud may be inappropriate.  However, accurate distance estimates require maser parallaxes with high accuracy (e.g., Hirota et al.~2008), which do not yet exist for the entire cloud. 
To simplify our calculations, we adopt a single fiducial distance of  250$\pm$50~pc,   
similar to other recent cloud-wide studies of Persues  \citep{eno06, jor06, rid06data, reb07, evans09}, 
 and caution that there may be significant differences in the distance  to different parts of the cloud.
   Perseus contains two rich protostellar clusters,  IC~348 and NGC~1333, and a number of other regions of active star formation 
  including  B5, B1, L1448, and L1455 (see Figure~\ref{perseus_map}).  Surveys in the infrared, sub-millimeter, and millimeter reveal a large population of low-mass pre-main sequence stars, embedded protostars, and starless cores in Perseus  
     \citep{lad93,aspin94,lad95, hat05, eno06, kirk06, muench07, guter08, evans09}. Only one B5 star in IC~348 (HD~281159) is confirmed to reside in the Perseus cloud, but there might be a few other high-mass stars that interact with the cloud 
    (through their winds and/or UV radiation)  
     even though they were not necessarily formed in the cloud complex (see, e.g., Walawender et al.~2004; Ridge et al.~2006b; Kirk et al.~2006; Rebull et al.~2007). 
   There is also a large number of nebulous objects associated with outflow shocks (i.e., HH objects and H$_2$ knots ) that have been identified in the cloud complex \citep{bal96a,bal97,yan98,wal05, davis08}.

The whole Perseus region was first surveyed in $^{12}$CO by \citet{sar79}, and since then has been mapped in CO at different angular resolutions (all with beams $>1\arcmin$) by a number of other authors \citep[e.g.,][]{bachiller86,ungerechts87,padoan99,sun06}. These maps show a clear velocity gradient in the Perseus molecular cloud
complex where the central cloud (LSR) velocity increases from about 4.5 km~s$^{-1}$ at the western edge of the cloud to about 10~km~s$^{-1}$ at the eastern end. The large velocity gradient in the gas across the entire complex 
and the fact that different parts of the Perseus cloud appear to have different distances (see above) could possibly indicate that the complex is made up of a superposition of different entities.   
Recently, the Perseus molecular cloud complex was also observed (and studied) in its entirety in the mid- and far-infrared as part of the 
``From Molecular Cores to Planet-forming Disks'' (ak.a., c2d) {\it Spitzer} Legacy Project \citep{jor06,reb07, evans09}.


\

\section{Data}
\label{datasection}

 In this paper, we use the $^{12}$CO(1-0) and $^{13}$CO(1-0) data collected for Perseus as part of the COMPLETE (COordinated Molecular Probe Line Extinction Thermal Emission) Survey of Star Forming Regions\footnote{see http://www.cfa.harvard.edu/COMPLETE}, described in detail by \citet{rid06data}.  
The $^{12}$CO and $^{13}$CO molecular line maps were observed between 2002 and 2005 using the 14-meter Five College Radio Astronomy Observatory (FCRAO) telescope with the SEQUOIA 32-element focal plane array.  The receiver was used with a digital correlator providing a total bandwidth of 25 MHz over 1024 channels.  The $^{12}$CO J=1-0 (115.271 GHz) and the $^{13}$CO J=1-0 (110.201 GHz) transitions were observed simultaneously using an on-the-fly  (OTF) mapping technique. The beam telescope at these frequencies
is about $46\arcsec$.  
 Both maps of $^{12}$CO and $^{13}$CO are essential for a thorough study of the outflow and cloud properties. The $^{12}$CO(1-0) is a good tracer of the cool and massive molecular outflows and provides the information needed to study the impact of these energetic phenomena on the cloud.  The $^{13}$CO(1-0) provides an estimate of the optical depth of the $^{12}$CO(1-0) line and can be used to probe the cloud structure and kinematics.

Observations were made in $10\arcmin \times 10\arcmin$  maps with
 an effective velocity resolution of 0.07~km~s$^{-1}$. These small maps were then patched together to form the final large map of Perseus, which  is about $6.25\arcdeg \times 3\arcdeg$.  Calibration was done via the chopper-wheel technique (Kutner \& Ulich 1981), yielding spectra with units of $T_A^*$. 
 We removed noisy pixels that were more than 3 times the average rms noise of the data cube, the entire map was then resampled to a $46\arcsec$ 
 grid, and the spectral axis was Hanning smoothed\footnote{see www.cfa.harvard.edu/COMPLETE/projects/outflows.html for a link to the molecular line maps} (necessary to keep the cubes to a size manageable by the 3D visualization code, see below). 
  During the observations of the Perseus cloud, different OFF positions were used depending on the location that was being mapped.  Some of these OFF positions had faint, though significant, emission which resulted in an artificial absorption feature
  in the final spectra.  Gaussians were fitted to the negative feature in regions with no gas emission, and the fits were then used to correct for the contaminating spectral component. The resulting mean $3-\sigma$ rms per channel in the $^{12}$CO and $^{13}$CO maps are 0.25 and 0.20~K, respectively, in the  $T_A^*$ scale.  
 Spectra were corrected for the main beam efficiencies of the telescope (0.49 and 0.45 at 110 and 115 GHz, respectively), obtained from 
 measurements of Jupiter.

\begin{figure}
\epsscale{1.0}
\plotone{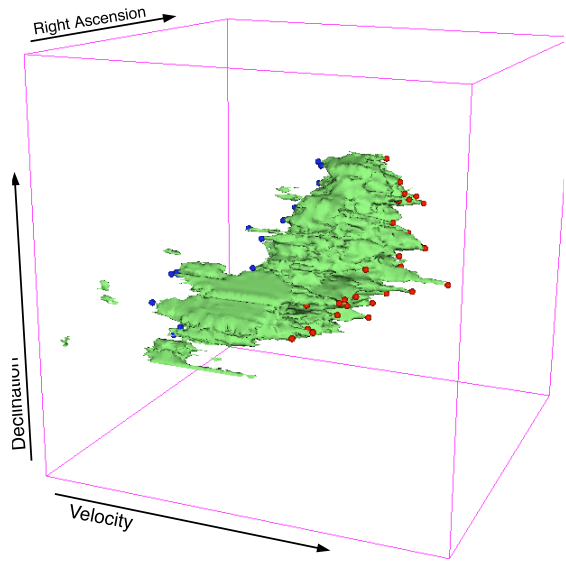}
\caption{Three-dimensional rendering of the molecular gas in B5 (i.e., Area VI in Figure~\ref{perseus_map}), using 3D Slicer. The green isosurface model shows the $^{12}$CO emission in 
 position-position-velocity space. The small circles show the locations of identified high-velocity points.
 \label{slicer}}
\end{figure}

\section{Computational Motivation and 3D Visualization}

This study  allows for a test of the effectiveness of 3D visualization of molecular line data of molecular clouds in RA-DEC-Velocity ({\it p-p-v}) space 
as a way to identify  velocity features, such as outflows, in large maps.\footnote{This work is done as part of the Astronomical Medicine  project (http://am.iic.harvard.edu) at the Initiative in Innovative Computing at Harvard (http://iic.harvard.edu).  The goal of the project is to address common research challenges to both the fields of medical imaging and astronomy including visualization, image analysis, and accessibility of large varying kinds of data.}   The primary program used for 3D visualization  is 3D Slicer\footnote{http://www.slicer.org/} which was developed originally at the MIT Artificial Intelligence Laboratory and the Surgical Planning Lab at Brigham and WomenÕs Hospital.  It was designed to help surgeons in image-guided surgery, to assist in pre-surgical preparation, to be used as a diagnostic tool, and to help in the field of brain research and visualization \citep{ger99}.  3D Slicer was first used with astronomical data by \citet{bor05} to study the hierarchical structure of star forming cores and velocity structure of IC~348 with $^{13}$CO(1-0) and C$^{18}$O(1-0) data.

We divided the Perseus cloud into six areas (with similar cloud central LSR velocities) for easier visualization and outflow search in 3D Slicer (see below).  The borders of these areas are similar to those named by \citet{pineda08}, who also  based their division mainly on the cloud's central LSR velocity. 
 The regions, whose outlines are shown in Figure \ref{perseus_map}, overlap between 1-3 arcminutes to guarantee complete analysis.  This overlap was checked to be sufficient based on the fact that new and known outflows which crossed regions were successfully double-identified.

For each area, an isosurface (constant intensity level) model was generated in 3D Slicer, using the  $^{12}$CO(1-0) map.  The threshold  emission intensity level chosen for each isosurface model was the lowest level of emission above the rms noise level for that particular region.  This creates a 3D model representing all of the detected emission.  The high velocity gas in this 3D space can be identified in the form of spikes, as shown for the B5 region in Figure~\ref{slicer}, which visually stick out from the general distribution of the gas.  These sharp protrusions occur since one is looking at the radial velocity component of the gas along the line of sight, thus causing spikes wherever there is gas at distinct  velocities far away from the main cloud velocity.  Instead of having to go through each region and carefully examine each channel map, or randomly scroll through the spectra by hand, this visualization allows one to instantly see where the high velocity points are located \citep[see also][]{bor07,bor08}.

\begin{figure*}
\epsscale{1.4}
\plotone{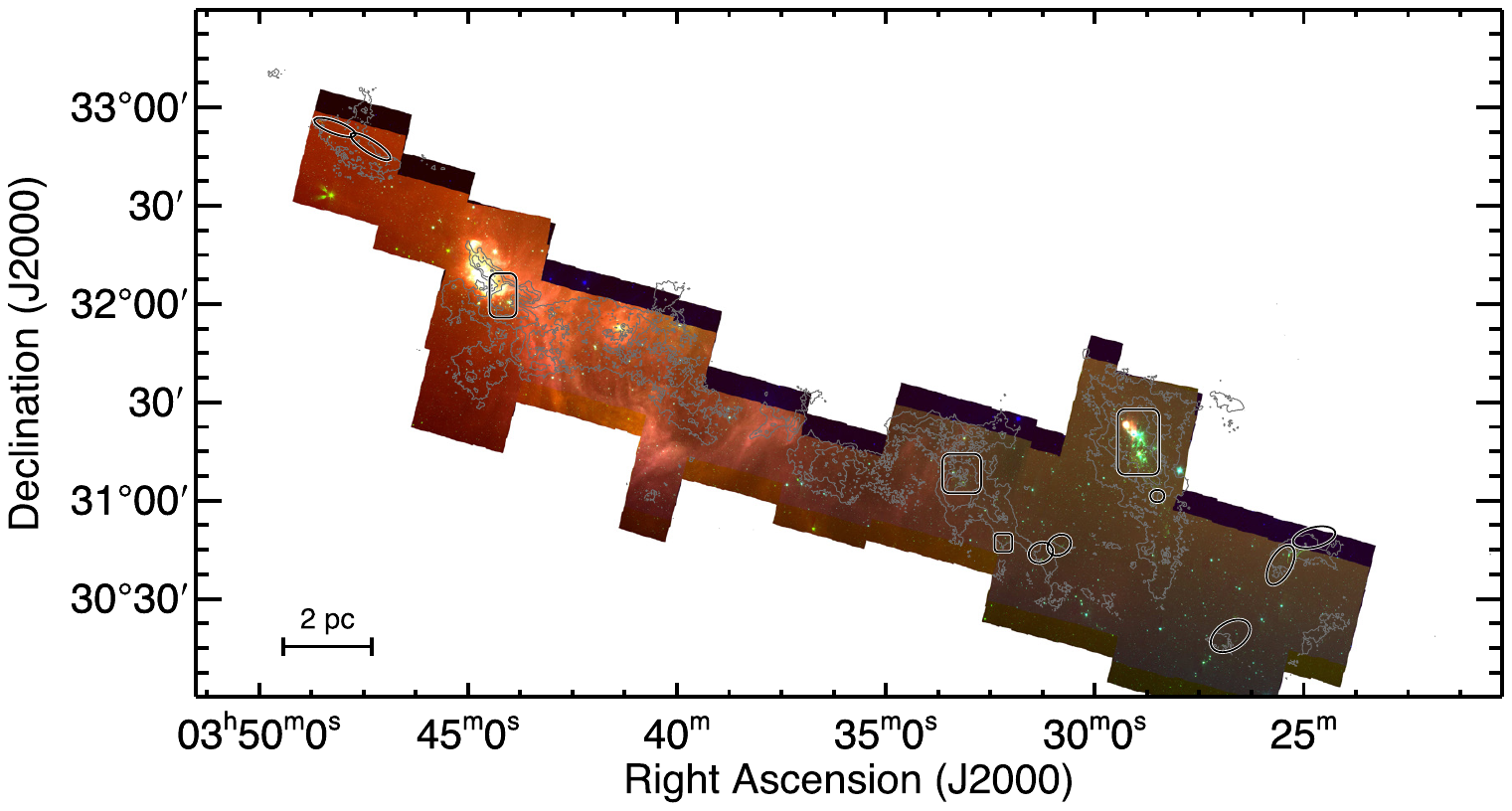}
\vspace{-4in}
\caption{{\it Spitzer} IRAC color image of the c2d coverage of the Perseus cloud (Evans et al.~2009). The color code is blue 
(3.6\micron), green (4.5\micron), and red (8.0 \micron). Ellipses and squares with rounded corners show the approximate regions where previously known outflows in Perseus lie. The grey contours show the 4 K~km~s$^{-1}$ level of the $^{13}$CO(1-0) integrated intensity map
(not corrected for the FCRAO beam efficiency). 
\label{knownflowsfig}}
\end{figure*}

\section{Outflow Identification}
\label{outid}

 A total of 218 high velocity points were visually identified in 3D Slicer for all of Perseus in $^{12}$CO. 
 We checked the position of each high-velocity point against the locations of known outflows (based on an extensive literature search) to determine if the point
 is associated with any known molecular outflow. From the 218 high-velocity points found, a total of 36 points were identified as associated with known molecular outflows. Figure~\ref{knownflowsfig} show the approximate regions where previously known $^{12}$CO(1-0) outflows lie.
 The number of high velocity points associated with a single outflow varies depending on its size and intensity.  For example, the parsec-scale B5 IRS1 outflow 
  is a conglomerate of 6 high velocity points whereas the  HH~211 outflow, which is only $\sim 0.1$ pc long, is identified by only one point. 
 We inspected each of the  remaining 182 high-velocity points
  to verify whether they are outflow-related or caused by other velocity features in the cloud.
  To determine if a high-velocity point is  outflow-related, we checked the spectrum by eye to look for outflow traits (e.g., high velocity low-intensity wings) and verified its proximity to known outflows and outflow sources \citep{wu04}, HH objects \citep{wal05},
  H$_2$ knots  \citep{davis08}, candidate young stellar objects (YSOs)
  form the c2d {\it Spitzer} survey \citep{evans09} and other known outflow sources and YSOs.  
  We also checked the velocity distribution and morphology of the gas associated with each high-velocity point to verify
 whether the velocity and structure of the gas were significantly different from that of the cloud in that region. 
 From the remaining 182 high-velocity points found, a total of 60 points were classified as being outflow {\it candidates}  based on the criteria mentioned above.  
 For 97\% of these outflow candidates, the maximum velocity away from the cloud velocity is equal to or greater than the escape velocity in that region of the cloud.
 We note that we purposely chose not to be too restrictive in the definition of outflow candidate 
 (e.g., we identified outflow candidates even without a solid outflow source identification, see below). 
 Using our broad, yet realistic, definition we can calculate the maximum possible impact from all plausible
 molecular outflows to the cloud.   
 Out of the remaining 122 points, 17 points were discarded due to too much noise or being pixels cut-off by the map's edge and 
 the other 105 points  are  thought to be caused by a number of other kinematic phenomena, including clouds at other velocities in the same line of sight unrelated to the Perseus cloud and
 spherical winds from young stars that produce expanding shell-like structures in the molecular gas 
 (as opposed to the discrete blob morphology observed in the 60 outflow candidates).     
 The distribution  and impact of these expanding shells on the cloud will be discussed further in a subsequent paper (Arce et al., in preparation)

\begin{deluxetable*}{lccccccp{4.2cm}}
\tabletypesize{\scriptsize}
\tablecaption{Candidate New and Extended Outflow Locations
\label{cpoctab}}
\tablewidth{0pt}
\tablehead{
\colhead{Name} & \colhead{RA} & \colhead{DEC} & \colhead{Area} &  
\colhead{Mass} & \colhead{Momentum} &  \colhead{Kinetic Energy}  & \colhead{Driving Source} \\
\colhead{ } & \multicolumn{2}{c}{(J2000)} & \colhead{(arcmin)} &  
 \colhead{(M$_{\sun}$)} & \colhead{(M$_{\sun}$ km s$^{-1}$)} & \colhead{(10$^{42}$ ergs)} &
 \colhead{Candidate(s)} 
}
\startdata

CPOC 1  & 03:23:21 & 30:52:10 & $19\times12$ &  0.05 & 0.19 &  6.93 & L1448-IRS1\\
CPOC 2  & 03:23:54 & 30:48:10 & $16\times7$  &  0.36 & 0.88 & 21.68 & L1448-IRS1\\
CPOC 3  & 03:24:30 & 30:50:00 & $10\times5$  &  0.02 & 0.08 &  2.93 & L1448-IRS3\\
CPOC 4  & 03:24:54 & 30:43:10 & $4\times4$   &  0.01 & 0.04 &  2.10 & multiple in L1448\\
CPOC 5  & 03:25:39 & 30:28:20 & $7\times5$   &  0.02 & 0.05 &  1.32 & SSTc2dJ032519.52+303424.2\\
CPOC 6  & 03:27:55 & 31:19:50 & $4\times3$   &  0.02 & 0.03 &  0.36 & multiple NGC1333, near HH338\\
CPOC 7  & 03:28:00 & 31:03:40 & $15\times12$ &  0.29 & 1.79 & 112.00 &  SSTc2dJ032834.49+310051.1  \\
CPOC 8  & 03:28:32 & 30:28:20 & $8\times11$  &  0.11 & 0.28 &  7.17 &  near HH750 and  HH743, SSTc2dJ032835.03+302009.9 or SSTc2dJ032906.05+303039.2\\
CPOC 9  & 03:28:28 & 31:13:20 & $8\times8$   &  0.26 & 0.56 & 12.63 & SSTc2dJ032832.56+311105.1 or SSTc2dJ032837.09+311330.8  \\
CPOC 10  & 03:28:27 & 31:23:20 & $8\times8$   &  0.24 & 0.42 &  7.50 & SSTc2dJ032844.09+312052.7  \\
CPOC 11  & 03:28:40 & 31:07:10 & $8\times6$   &  0.11 & 0.27 &  7.01 &  STTc2dJ032834.53+310705.5 \\
CPOC 12  & 03:28:43 & 31:07:30 & $8\times7$   &  0.19 & 0.97 & 52.02 &  SSTc2dJ032843.24+311042.7  \\
CPOC 13  & 03:28:50 & 31:27:10 & $6\times8$   &  0.31 & 0.80 & 21.00 & multiple in NGC1333  \\
CPOC 14  & 03:28:57 & 30:50:20 & $6\times5$   & 0.03  & 0.05 &  0.73 & SSTc2dJ032850.62+304244.7 or SSTc2dJ032852.17+304505.5\\
CPOC 15  & 03:29:07 & 30:45:50 & $7\times5$   & 0.19  & 0.80 & 32.82 & SSTc2dJ032850.62+304244.7 or SSTc2dJ032852.17+304505.5\\
CPOC 16  & 03:29:30 & 31:07:10 & $6\times6$   &  0.04 & 0.10 &   2.40 & HH18A, multiple in NGC1333       \\
CPOC 17  & 03:29:41 & 31:17:30 & $9\times13$  &  3.20 & 8.49 & 235.28 & near HH497, HH336, multiple in NGC1333\\ 
CPOC 18  & 03:29:41 & 31:27:10 & $5\times6$   &  0.08 & 0.21 & 6.35   & HH764, multiple in NGC1333\\
CPOC 19 & 03:29:27 & 31:34:00 & $9\times7$   & 0.19  & 0.59 & 19.31 & IRAS03262+3123\\
CPOC 20 & 03:30:06 & 31:27:10 & $5\times4$   & 0.04  & 0.08 &  1.73 & multiple NGC1333\\
CPOC 21 &  03:30:11 & 31:14:00 & $8\times5$   & 0.05  & 0.13 &  3.45 & HH767, SSTc2dJ033024.08+311404.4 \\ 
CPOC 22 & 03:30:40 & 30:37:00 & $6\times11$  & 0.30  & 1.07 & 39.24 & multiple in Per6 aggregate\\
CPOC 23 & 03:30:56 & 31:21:10 & $6\times6$   & 0.01  & 0.05 &  3.56 & multiple in NGC1333 or B1\\
CPOC 24 & 03:31:23 & 31:01:30 & $27\times18$ &  0.46  & 2.99 &  193.71 & multiple in B1 or B1-Ridge\\
CPOC 25 & 03:31:23 & 31:20:40 & $4\times7$   & 0.02  & 0.14 &  9.73 & multiple in NGC1333 or B1\\
CPOC 26 & 03:31:40 & 30:54:40 & $6\times4$   & 0.09  & 0.27 &  8.26 & IRAS03292+3039 or others in B1 and B1-Ridge\\
CPOC 27 & 03:31:54 & 31:14:10 & $8\times5$   & 0.07  & 0.40 & 21.85 & multiple in NGC133 or B1\\
CPOC 28 & 03:32:04 & 30:40:20 & $4\times5$   & 0.58  & 1.35 & 31.94 & multiple in B1-Ridge\\
CPOC 29 & 03:32:25 & 31:18:10 & $5\times7$   & 0.06  & 0.17 &  4.89 & multiple in B1\\
CPOC 30 & 03:32:37 & 31:02:50 & $3\times6$   & 0.04  & 0.13 &  3.57 & multiple in B1\\
CPOC 31 & 03:32:58 & 31:22:20 & $4\times8$   & 0.15  & 0.37 &  9.32 & SSTc2dJ033312.84+312124.2 or SSTc2dJ033313.80+312005.3\\
CPOC 32 & 03:33:14 & 30:59:00 & $4\times6$   & 0.07  & 0.17 &  4.25 & SSTc2dJ033346.92+305350.1 or multiple in B1 core\\
CPOC 33 & 03:33:40 & 31:28:50 & $5\times6$   & 0.21  & 0.50 &  11.77 & multiple in B1\\
CPOC 34 & 03:33:58 & 31:16:10 & $6\times4$   & 0.14  & 0.25 &  4.58 & SSTc2dJ033401.66+311439.8\\
CPOC 35 & 03:34:43 & 31:22:00 & $5\times8$   & 0.10  & 0.24 &  5.43 & SSTc2dJ033430.78+311324.4 or SSTc2dJ033449.84+311550.3\\
CPOC 36 & 03:35:10 & 31:18:00 & $4\times5$   & 0.11 & 0.30 &  8.02 & SSTc2dJ033430.78+311324.4 or SSTc2dJ033449.84+311550.3\\
CPOC 37 & 03:38:58 & 32:05:50 & $5\times3$   & 0.16  & 0.19 &  2.34 & unknown between IC348 and B1\\
CPOC 38 & 03:39:05 & 32:08:40 & $5\times4$   & 0.04  & 0.06 &  0.90 & unknown between IC348 and B1\\
CPOC 39 & 03:39:11 & 31:19:00 & $8\times8$   & 0.10  & 0.19 &  3.83 & SSTc2dJ033915.81+312430.7 or SSTc2dJ034001.49+311017.3\\
CPOC 40 & 03:39:16 & 32:18:10 & $6\times4$   & 0.09  & 0.24 & 11.17 & IRAS03363+3207\\
CPOC 41 & 03:39:18 & 31:58:10 & $7\times6$   & 0.14  & 0.21 &  3.22 & IRAS03367+3147\\
CPOC 42 & 03:39:20 & 32:17:40 & $5\times5$   & 0.20  & 0.19 &  1.92 & IRAS03363+3207\\
CPOC 43 & 03:40:24 & 32:04:00 & $7\times8$   & 2.04  & 3.89 & 76.37 &  IRAS03367+3147 or multiple west of IC348\\ %
CPOC 44 & 03:42:12 & 31:51:50 & $5\times5$   & 0.09  & 0.18 &  3.44 & multiple east of IC348\\
CPOC 45 & 03:44:34 & 31:58:20 & $4\times6$   & 0.54  & 0.73 &  10.04 & multiple in south edge of IC348\\
CPOC 46 & 03:44:53 & 32:14:40 & $11\times6$  & 0.33  & 0.66 &  13.19 & multiple in north edge of IC348\\
CPOC 47 & 03:44:58 & 32:32:00 & $11\times9$  & 0.20  & 0.41 &  8.69 & multiple in north edge of IC348\\
CPOC 48 & 03:45:01 & 31:57:50 & $4\times4$   & 0.16  & 0.25 &  3.94 & multiple in south edge of IC348\\
CPOC 49 & 03:45:04 & 32:00:30 & $5\times5$   & 0.03  & 0.09 &  3.40 & multiple in south edge of IC348\\
CPOC 50 & 03:45:26 & 31:58:00 & $6\times6$   & 0.25  & 0.36 &  5.29 & multiple in south edge of IC348\\
CPOC 51 & 03:45:53 & 32:34:00 & $7\times7$   & 0.27  & 0.32 &  4.09 & B5-IRS1\\
CPOC 52 & 03:45:59 & 32:42:50 & $7\times7$   & 0.13  & 0.29 &  6.55 & unknown in B5\\
CPOC 53 & 03:46:54 & 32:36:20 & $6\times5$   & 0.07  & 0.10 &  1.24 & B5-IRS3\\
CPOC 54 & 03:47:16 & 32:39:50 & $5\times4$   & 0.06  & 0.09 &  1.48 & B5-IRS3\\
CPOC 55 & 03:47:17 & 33:01:40 & $15\times15$ & 3.97  & 7.55 & 147.19 & B5-IRS4?\\
CPOC 56 & 03:47:60 & 32:38:40 & $20\times13$ & 3.76  & 6.73 & 124.04 & multiple in B5\\
CPOC 57 & 03:48:01 & 33:14:40 & $6\times6$   & 0.09  & 0.11 &  1.45 & B5-IRS4?\\
CPOC 58 & 03:49:14 & 32:57:40 & $7\times6$   & 0.28  & 0.27 &  2.83 & unknown in B5\\
CPOC 59 & 03:49:18 & 33:04:40 & $5\times7$   & 0.20  & 0.25 &  3.37 & B5-IRS1\\
CPOC 60 & 03:49:41 & 33:12:20 & $8\times7$   & 0.58  & 0.64 &  7.08 & unknown in B5\\

\enddata

\end{deluxetable*}

\begin{figure}
\epsscale{1.0}
\plotone{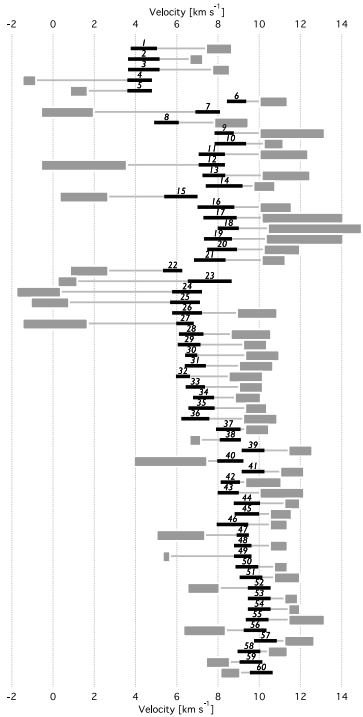}
\caption{Velocity of CPOCs with respect to the local cloud central (LSR) velocity. Black (thick) horizontal lines are centered on the CPOC area's central LSR velocity and their extent represent the velocity dispersion of the cloud, as measured from the average $^{13}$CO spectra over the CPOC area (see Table~\ref{cpoctab}). The grey horizontal rectangles indicate the range of velocities for each CPOC. Numbers above each black line refer to the CPOC number. 
 \label{cpoc_vel}}
\end{figure}

  We visually inspected the velocity maps in the area surrounding each of the 60 high-velocity points identified as 
 outflow in origin  (but unrelated to known outflows) and chose
 an area (in RA-Dec space) and velocity range that included all or most of the emission associated with the kinetic feature.
The integration area and velocity ranges were conservatively chosen to include only the emission visibly associated with 
the outflowing material, thus avoiding cloud emission.  The high-velocity gas associated with these 60 points show discrete morphologies in area and velocity.
Hereafter each of these high-velocity features is referred as a ``COMPLETE Perseus Outflow Candidate" (CPOC) and we list their positions and other properties 
 in Table~\ref{cpoctab}\footnote{see www.cfa.harvard.edu/COMPLETE/projects/outflows.html for a link to the fits cubes and the integrated intensity fits files of the CPOCs, as well as a list of the YSO candidates, HH objects and H$_2$ knots in the cloud}. 
 In Figure~\ref{cpoc_vel} we show the velocity ranges of all CPOCs, in comparison with their local cloud (LSR) velocity.
 
Our outflow-detection technique proved to be reliable, as we detect high-velocity gas associated with all published CO(1-0) outflows
(see Figure~\ref{knownflowsfig}). However,
it is very probable that the catalog of new molecular outflows generated for this paper is an underestimate of the true number of previously undetected molecular
outflows due to the resolution of the CO maps and other limitations of our outflow-detection technique.
Unknown outflows that are smaller than the beam size of our map (i.e., $0.06$~pc at the assumed distance of Perseus) or that have weak high-velocity wings (i.e., with intensities less than twice the rms of the spectra at that particular position) cannot not be detected by our technique. Outflows with maximum velocities too close to the ambient gas velocity to produce a detectable high-velocity spike in $p-p-v$ space are also missed by our procedure, as well as high-velocity gas in regions ``contaminated'' by unrelated clouds along the same line-of-sight. 
Although we are able to detect high-velocity outflow gas in regions with a high density of protostars,
our map's beam size limits our ability to distinguish individual outflows in dense clusters like NGC~1333. 
Higher resolution maps would be needed to identify the individual molecular outflows in regions with a high-density of protostars.  
However, the technique used in this paper proved ideal for finding parsec-scale outflows and identifying outflows over large areas.

\begin{figure*}
\epsscale{1.1}
\plotone{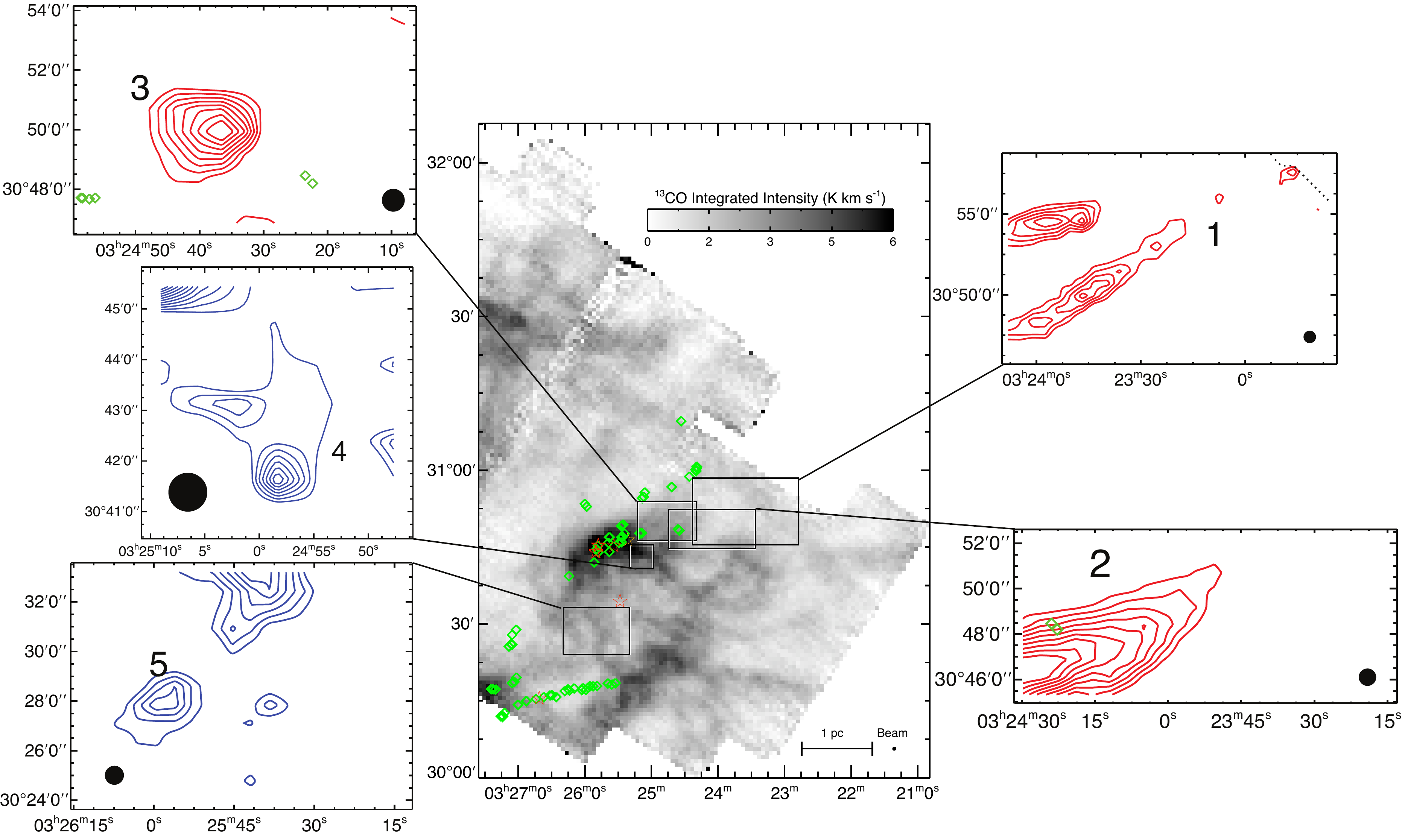}
\caption{CPOCs in Area I. The grey-scale image shows the $^{13}$CO integrated intensity. Star symbols indicate the position of candidate YSOs from the c2d survey and known outflow and IRAS sources, while diamonds represent HH objects and H$_2$  outflow shock emission. Integrated intensity contour maps of the CPOCs in Area I are shown. Redshifted (blueshifted) CPOCs are shown in black (grey) contours. The velocity range of integration is that shown in Figure~\ref{cpoc_vel}. For all panels the x-axis shows the Right Ascension (J2000), and the y-axis shows the Declination (J2000). 
The telescope beam is shown as a filled black circle in the lower part of each CPOC map. The $^{13}$CO map is not corrected for the FCRAO beam efficiency. 
\label{fig_reg1}}
\end{figure*}

Although most  molecular outflows in Perseus are in well-known  regions of active star formation (i.e, L1448, NGC~1333, B1, IC~348, and B5) we also detected high-velocity features associated with molecular outflows across the whole expanse of the cloud complex. The locations of all the CPOCs can be seen in Figures~\ref{fig_reg1} to \ref{fig_reg6}. From these figures it is clear that most of the {\it new} outflow candidates are located around the edges of clusters (e.g., CPOCs 16 to 22 around NGC~1333, and CPOCs 48 to 50 in the outskirts of the IC~348 cluster), or were identified  in  cloud regions between clusters (e.g., CPOCs 38 to 42). 
 By surveying the entire cloud complex and regions between clusters, we were able to identify previously-undetected outflows 
and extensions of known outflows in the poorly-studied regions of the Perseus cloud complex.
We discuss all candidate outflows and their possible association with nearby protostellar sources, HH objects and H$_2$ knots in the Appendix.

 \subsection{Outflow Source and Counter-Lobe Identification}
 \label{sourceid}

  For many of the new outflow candidates it is hard to unambiguously assign a source. Pervious studies show that not all molecular outflows have lobes that originate close to the source and extend with a continuous structure all the way to the outflow's  terminus.
Instead, many molecular outflows are composed of discrete high-velocity blobs along their axes produced by different mass ejection episodes (e.g., Cernicharo \& Reipurth 1996; Yu et al.~1999; Arce \& Goodman 2001a; Arce \& Goodman 2002b). In some cases the high-velocity CO blobs coincide with one or more HH objects, while in other cases there are no shock tracers 
in the vicinity of the outlowing gas. 
In addition, a number of known processes including  precession \citep{terquem99}, a relative motion between the outflow source and the ambient medium (Bally \& Reipurth 2001; Masciadri \& Raga 2001; Goodman \& Arce 2004), and collision with a denser environment \citep{raga02} may cause the axis of an outflow to change over time. 
The episodic nature of outflows as well as possible variations in their axes  make the reliable 
identification of an outflowing blob's source difficult, especially in regions with a dense population of young stellar objects. 
It is, therefore, no surprise that our data does not allow us to determine for certain the sources associated with the listed CPOCs. Nonetheless, 
in Table~\ref{cpoctab} we give possible candidate sources based on their proximity to the CPOCs.
In addition to the factors listed above, the fact that sometimes molecular outflows have one detectable lobe ---either due to contamination from another cloud along the same line of sight or the fact that the other lobe breaks out from the molecular cloud---
makes it very difficult to pair each CPOC with a corresponding counter lobe.
 Higher angular resolution CO maps in combination with spectroscopic and proper motion studies (to determine the three-dimensional velocity) of nearby HH objects should allow the proper identification of the source and coounter-lobe for most CPOCs.

 \subsection{Outflow Candidates vs. Turbulence Features}

Despite the fact that outflow-related high-velocity features could in principle be confused with
turbulent-generated velocity structures, we are certain that most (if not all) CPOCs are outflow in origin.
Numerical simulations show that line profiles from a medium with a random turbulent velocity field, such as a molecular cloud, will exhibit non-Gaussian features like double peaks, skewness, and high-velocity wings, even if generated using a Gaussian random field  (Dubinski et al.~1995). This could lead to confusion between high-velocity turbulent-generated velocity features in the molecular line and high-velocity outflow-generated velocity features. However, observational 
studies show that at  scales of a few tenths of parsecs (and densities of about $10^4$ cm$^{-3}$),
most  molecular clouds  exhibit 
  turbulence-related velocity features with maximum velocities of about 
1~km~s$^{-1}$ and all are less than 2~km~s$^{-1}$ away from the central cloud velocity (Falgarone et al.~1998). 
Moreover,  observations of clouds that are presumably free of star formation activity and where the only ``high'' velocity features should be caused by turbulence (or other clouds along the line of sight) the range in velocities of the emission is not more 
than 4~km~s$^{-1}$ (see  Falgarone et al.~1990; 1998; 2006; 2009). This implies that the highest velocity turbulence features are never more than 2~km~s$^{-1}$ away from the line center. 
In contrast, most (i.e., 97\%) of the candidate outflows listed in Table~1 have a maximum velocity that is greater than or equal to 2~km~s$^{-1}$, and all have a maximum velocity that is 1.9~km~s$^{-1}$ or more.  In addition,
 all CPOCs have a maximum velocity that is 3.1 times (or more) than the cloud's velocity dispersion ($\sigma_{cl}$)
 and 73\% of the CPOCs have maximum velocities of more than 4.5~$\sigma_{cl}$
  (see Figure~\ref{cpoc_vel}). The velocities of turbulent generated features in terms of the velocity dispersion of the cloud
are typically much lower than those exhibited by the CPOCs (see, e.g., Falgarone et al.~1990).
The CPOCs' high velocities, as well as their proximity to YSOs and HH objects makes it very likely that most (if not all) are   outflow-related features, and not produced by pure random motions in the cloud's turbulent gas.

\section{Analysis \& Discussion}
\label{analysis}

\subsection{Mass, Momentum and Energy of Outflows}
\label{outmass}

We use a method to obtain the outflow mass similar to that  described by Arce \& Goodman (2001, hereafter AG01). This technique, which is based on the method employed by Bally et al.~(1999) and Yu et al.~(1999), uses the  $^{12}$CO(1-0) to $^{13}$CO(1-0)  ratio to estimate the opacity in the $^{12}$CO(1-0)  line, as a function of velocity. In molecular clouds the $^{12}$CO(1-0) line
is typically optically thick and its opacity depends on velocity. In the general, at high outflow velocities the opacity of the line is lower than the opacity of the line close to the cloud velocity. Using an optically thick line without properly correcting for its velocity-dependent opacity will result in an underestimation of the outflow mass, momentum, and kinetic energy.

We briefly describe the method here, but for more detail see AG01.
For each candidate (or known) outflow we calculate average spectra of $^{12}$CO(1-0) and $^{13}$CO(1-0) over the defined outflow region (shown in Tables 1 and 2)
in order to estimate the ratio of  $^{12}$CO (1-0) to  $^{13}$CO (1-0) as a function of velocity. 
The line ratios, hereafter denoted $R_{12/13}$, were each fitted with a second order
polynomial as described in AG01.
To calculate the outflow mass at a given position,
we directly use the  $^{13}$CO emission at low outflow velocities 
(given by the main beam corrected antenna temperature of the line, $T_{mb}^{13}$).
At high outflow velocities, where the $^{13}$CO was not reliably detected, we use the  $^{12}$CO (1-0) emission ($T_{mb}^{12}$)
and the fit to  $R_{12/13}(v)$ to estimate the value of $T_{mb}^{13}$
at the given velocity and position, using the simple equation
$T_{mb}^{13} = T_{mb}^{12}/ R_{12/13}(v)$.
 We then obtain a value of the $^{13}$CO opacity ($\tau_{13}$),
 from which we then obtain a value of the $^{13}$CO column density ($N_{13}$) and then the mass, using
equations (1), (3), and (4) of AG01. We only use spectral data that is greater than or equal to three times the rms noise of the spectrum.

\begin{deluxetable*}{lccccccc}
\tabletypesize{\footnotesize}
\tablecaption{Properties of Known Outflows and Outflow Regions 
\label{outflowknowntab}}
\tablewidth{0pt}
\tablehead{
\colhead{Name} & \colhead{RA} & \colhead{DEC} & \colhead{Area} & \colhead{Velocity Range} & 
\colhead{Mass} & \colhead{Momentum} &  \colhead{Kinetic Energy}\\
\colhead{ } & \multicolumn{2}{c}{(J2000)} & \colhead{(arcmin)} & \colhead{(km s$^{-1}$)} &
 \colhead{(M$_{\sun}$)} & \colhead{(M$_{\sun}$ km s$^{-1}$)} & \colhead{(10$^{42}$ ergs)}
}
\startdata

L1448 Outflows (blue)         & 03:24:32 & 30:50:10 & $12\times6$   &  [-5.0, -0.5] & 0.06 &  0.34  &  21.0\\
L1448 Outflows (red)          & 03:25:25 & 30:40:30 & $8\times11$   &  [7.0, 16.0]  & 0.45 &  1.50  &  61.2\\
L1455 (blue)                  & 03:26:40 & 30:18:50 & $12\times8$   &  [0.1, 2.5]   & 0.10 &  0.27  &   7.8\\
L1455 (red)                   & 03:26:40 & 30:18:50 & $12\times8$   &  [7.6, 9.0]   & 0.11 &  0.34  &  11.0\\
IRAS 03262+3123 (red)         & 03:28:17 & 31:01:30 & $4\times3$    &  [9.5, 13.8]  & 0.08 &   0.29  &  10.7\\ 
NGC~1333 Outflows (red)       & 03:29:00 & 31:17:10 & $12\times14$  & [10.2, 12.3]  & 1.35 &  4.22 & 142.5\\
                                                      & 03:29:15 & 31:25:00 & $5\times8$    & [10.2, 19.6]  & 0.49 &  1.18 &  28.7\\
NGC~1333 Outflows (blue)     & 03:29:03 & 31:16:20 & $17\times14$  & [-0.6, 3.8]   & 0.50 &  2.62 & 142.0\\
                                                     & 03:29:20 & 31:27:00 & $12\times8$   & [-0.6, 0.8]   & 0.03 &  0.26 &  20.0\\
IRAS 03282+3035 (red)         & 03:31:09 & 30:47:00 & $6\times5$    &  [8.6, 11.8]  & 0.21 &  0.45  &   9.7\\
IRAS 03282+3035 (blue)        & 03:31:16 & 30:45:00 & $9\times6$    &  [-2.5, 1.2]  & 0.03 &  0.24  &  17.0\\
IRAS 03291+3039 (red)         & 03:32:18 & 30:47:30 & $5\times5$    &  [8.9, 11.2]  & 0.02 &  0.06  &   1.7\\
B1 Outflows (blue)            & 03:33:14 & 31:10:20 & $12\times12$  &  [-1.6, 1.0]  & 0.05 &  0.37  &  24.6\\
B1 Outflows (red)             & 03:33:14 & 31:10:20 & $12\times12$  &  [9.1, 12.5]  & 0.31 &  0.89  &  27.5\\
IC 348 Outflows (blue)        & 03:44:01 & 32:02:40 & $8\times13$   &  [5.8, 6.5]   & 0.79 &  1.89  &  45.4\\
IC 348 Outflows (red)         & 03:44:01 & 32:02:40 & $8\times13$   &  [10.3, 11.6] & 2.05 &  3.73  &  68.8\\
B5-IRS1 (blue)                & 03:48:39 & 33:01:00 & $9\times6$    &  [6.5, 8.6]   & 0.77 &  1.36  &  25.2\\
                              & 03:48:12 & 32:55:10 & $8\times4$    &  [5.8, 8.6]   & 0.93 &  1.79  &  36.3\\   
                              & 03:46:44 & 32:46:50 & $8\times7$    &  [7.1, 8.6]   & 0.35 &  0.67  &  13.1\\   
B5-IRS1 (red)                 & 03:48:26 & 32:57:30 & $7\times5$    &  [11.4, 12.8] & 0.22 &  0.37  &   6.4\\ 
                              & 03:47:17 & 32:50:40 & $12\times5$   &  [11.5, 14.4] & 0.55 &  0.91  &  15.9\\ 
                              & 03:46:46 & 32:44:30 & $9\times5$    &  [11.5, 14.4] & 0.58 &  0.87  &  13.8\\ 


\enddata
\end{deluxetable*}

 We estimate the excitation temperature, $T_{ex}$, for each CPOC assuming that the $^{12}$CO(1-0) line core is optically thick. We  
 measured the peak temperature of each spectrum in the CPOC region, and use Equation~2 in AG01  to obtain a distribution of  the ambient cloud $T_{ex}$ values within that area. 
 We obtain an average $T_{ex}$ for each CPOC region, and assume that this average value is the temperature  
 of the outflowing gas (at all velocities) in the entire CPOC region.   
Pineda et al.~(2008) found slight variations (from 2.8 to 4.9 $\times 10^5$)
 in the ratio of molecular hydrogen to $^{13}$CO  depending on the region in Perseus. We used  their region-dependent values of  [H$_2$]/[$^{13}$CO]  and  
 the ratio of   $^{12}$CO to $^{13}$CO of 62 (from Langer \& Penzias 1993).
 We obtain the line-of-sight outflow momentum, $P_{out}$, using 
\begin{equation}
P_{out} = \sum_{v} M(v_{out}) v_{out}
\end{equation}
where $v_{out}$ is the line-of-sight component of the outflow velocity and $M(v_{out})$ is the outflow mass as a function of (line-of-sight) outflow velocity.  The outflow velocity is defined as $v_{out} = v_{obs} - v_{amb}$, where $v_{obs}$ and $v_{amb}$ are observed and ambient cloud LSR velocities, respectively. For each CPOC we obtain a value of $v_{amb}$ (i.e., the line core velocity) by fitting a Gaussian to the average $^{13}$CO(1-0) spectrum over the CPOC area. The outflow kinetic energy, $E_{out}$, using only the line-of-sight component of the velocity is obtained with:
\begin{equation}
E_{out} = 0.5 \sum_{v} M(v_{out}) v_{out}^2
\end{equation}

\begin{figure*}
\epsscale{1.15}
\plotone{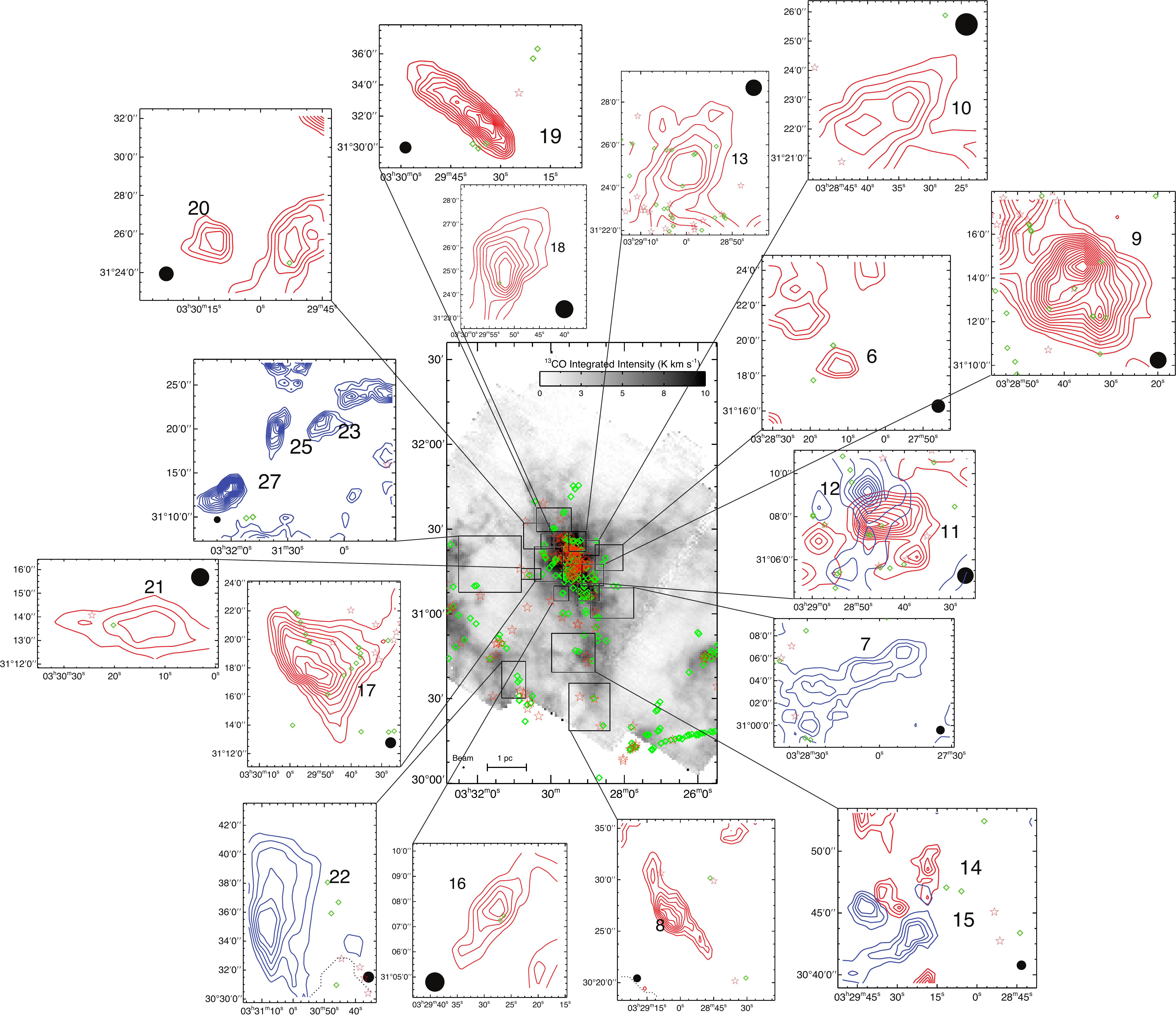}
\caption{CPOCs in Area II. The rest is the same  as in Figure~\ref{fig_reg1}
\label{fig_reg2}}
\end{figure*}

The numbers shown in Tables \ref{cpoctab} and \ref{outflowknowntab} represent lower limits, as there are several outflow properties that have not been taken into consideration when estimating the outflow mass, momentum, and energy which increase the total estimates.  As discussed in Section~\ref{outid},  we conservatively define the lowest outflow velocity as the velocity for which we are certain that most (or all) of the emission within the defined area arises from the outflow (i.e., there is little or no cloud emission in the region).  This results in an under-estimation of the outflow mass (as well as energy and momentum), as we are not including the outflow emission ``hidden'' under the cloud line emission.  Previous outflow studies indicate that failing to account for this hidden component will underestimate the mass by at least a factor of two (as shown by, e.g.,  Margulis \& Lada 1985).  

The assumed value of the outflow temperature can also affect the outflow mass estimates. We use the average excitation temperature of the cloud in the outflow region to estimate the outflow mass, as it is the only estimate of the gas temperature we can obtain with our data.  Outflow studies that observe different transitions of the same molecule show that temperatures of outflowing molecular gas are higher than the cloud's excitation temperature and is not uncommon for outflows to reach temperatures of 50 to 100~K (e.g., Hirano \& Taniguchi 2001; van Kempen et al.~2009).  When using the $^{12}$CO(1-0) emission to derive outflow properties, 
the dependence of CO column density with excitation temperature is such that an increase in the temperature above 10~K will result in an increase in the estimate of the outflow mass (see Appendix in Lada \& Fich 1996). 
If we assume that typical outflow temperatures are about 3 times higher than the cloud excitation temperature, the resulting estimate of the outflow mass, momentum, and energy would increase by a factor of about 2.5. 
Combing this factor and a factor of two to correct for the unaccounted outflow emission hidden under the cloud emission, we reckon that our original estimates of the outflow mass should increase by a factor of 5.

The values of the outflow momentum and energy shown in Tables \ref{cpoctab} and \ref{outflowknowntab} only take into consideration the line-of-sight outflow velocity component.  If we assume an average outflow inclination of $45\arcdeg$, the momentum and energy estimates increase by factors of 1.4 and 2, respectively. 
Also, some of the shocks that produce these outflows may be dissociative (see Reipurth \& Bally 2001), and thus in some outflows a fraction of the outflow momentum and energy may
 not be traceable by the CO emission. Combining these factors with the correction factor for the outflow mass, we estimate that  to obtain reasonable values of the outflows' momentum, and kinetic energy our original estimates should increase by ``correction'' factors of 7 and 10, respectively.

\begin{figure*}
\epsscale{1.0}
\plotone{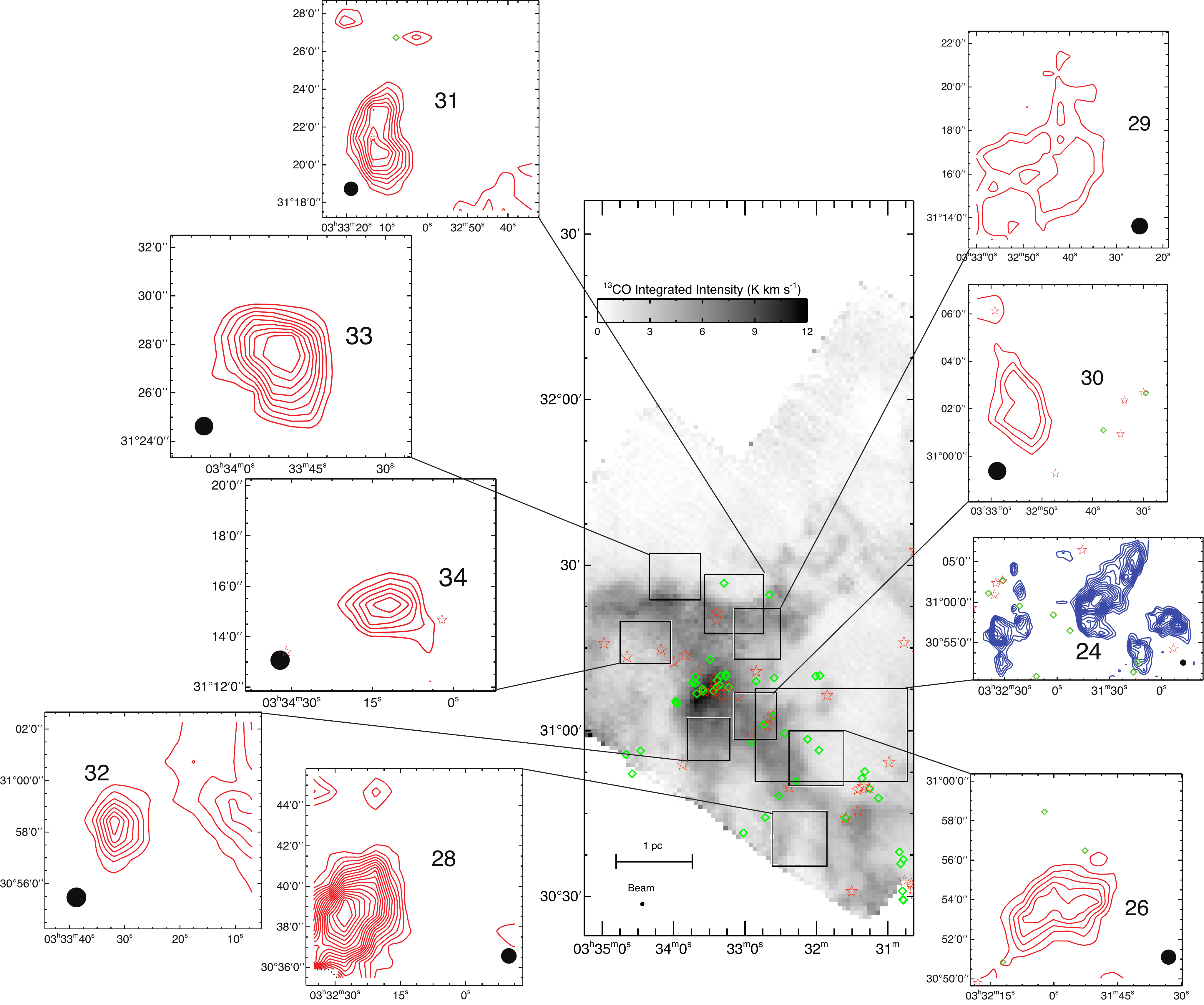}
\caption{CPOCs in Area III. The rest is the same  as in Figure~\ref{fig_reg1}.
\label{fig_reg3}}
\end{figure*}


%

The total mass, momentum and energy of all outflows in our maps (i.e., CPOCs and previously known outflows) are
163 M$_{\sun}$, 517 M$_{\sun}$~km~s$^{-1}$, and    $20.5 \times 10^{45}$ erg, respectively (using the correction factors discussed above). If we only include the numbers for the previously known outflows, the totals are 
51 M$_{\sun}$, 173 M$_{\sun}$~km~s$^{-1}$, and    $7.5 \times 10^{45}$ erg (see Table~\ref{alloutflowstab}).
Our study shows that in Perseus there is much more outflowing mass and considerably more injection of momentum and energy by outflows into the cloud than previously thought, as the 
 new CPOCs more than double the total outflow mass, momentum, and  kinetic energy in the
 Perseus molecular cloud complex.
 It is clear that large-scale observations like the ones used in this study are necessary to obtain a complete picture of the impact of outflows on their entire host cloud.

\begin{figure*}
\epsscale{1.0}
\plotone{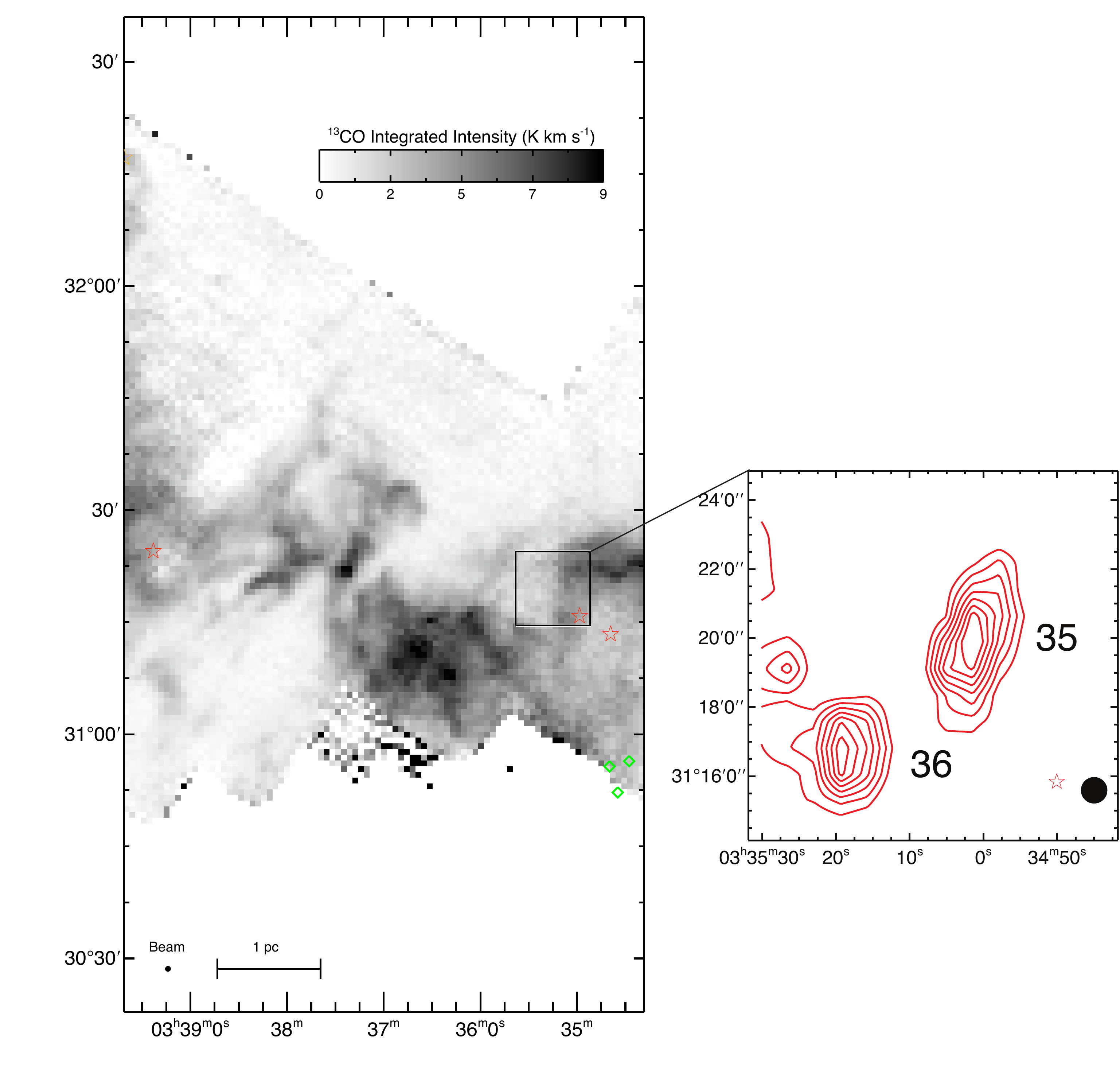}
\caption{CPOCs in Area IV. The rest is the same  as in Figure~\ref{fig_reg1}.
\label{fig_reg4}}
\end{figure*}

\subsection{The Impact of Outflows in Perseus}

With our cloud-wide survey we can assess the impact of outflows on the Perseus complex as a whole by comparing the total energy and momentum of all the observed outflows with the cloud complex's energetics.  From the COMPLETE molecular line data (and using the procedure described above), we obtain a mass for the observed  Perseus cloud complex of approximately  
$7 \times 10^3$ M$_{\sun}$. The
average velocity width (FWHM) is about 2~km~s$^{-1}$, so the turbulent energy of the complex is  
$E_{turb} \sim 1.6 \times 10^{47}$ erg.  The total kinetic energy from all the observed outflows (using the
correction factor discussed above) is $2 \times 10^{46}$ ergs, only 13\% of the turbulent energy of the entire complex.
Even with the considerable increase in the outflow energy and momentum injection with the new outflow candidates reported here,
it  is evident that the energy input solely by protostellar outflows is not enough to feed the observed turbulence in the cloud.  This should not come as a surprise, as although we find numerous outflows in the cloud they are mostly found in concentrated regions of star formation and there are large extents of molecular gas in the complex with few or no outflows. This implies that an additional energy source is responsible for turbulence on a global cloud scale in the Perseus complex. In a subsequent paper we propose that 
expanding shells produced by spherical winds (and soft-UV radiation) from stars in, and near, the cloud can  provide the additional energy required to drive and maintain the turbulence throughout the entire cloud complex (Arce et al., in preparation).

The clustering of outflows in localized star forming regions indicates the possibility that although the outflows might only have a small impact on the Perseus complex as a whole, they still may have considerable impact on their immediate environment.  We define six regions of active star formation in the Perseus molecular cloud complex where we find a cluster or group of outflows:
L1448, NGC~1333, B1-ridge, B1, IC~348, and B5.  The locations of these regions within the cloud complex and their extent are shown in Figure~\ref{perseus_map} and Table~\ref{regiontab}, their physical properties are shown in Table~\ref{physregtab}, and the total outflow mass, momentum and energy within these regions are shown in Table~\ref{outregtab}.  In order to quantitatively assess the impact of outflows on their local environment, we compare the outflow energy and momentum with their host region's energetics.

\subsubsection{Outflows and Turbulence}
\label{outturb}
A comparison between the total outflow kinetic energy in each region and the region's turbulent energy shows that the total energy of outflows is between about 14 and 80\% of the total turbulent energy of the region (see Table~\ref{comptab}).  This suggests that in some regions, outflows in a localized area of star formation inside a cloud complex can inject a significant amount of energy into the gas to considerably affect the turbulence of the local environment.  

\begin{deluxetable}{lccc}[h!]
\tabletypesize{\small}
\tablecaption{Perseus Outflow Properties
\label{alloutflowstab}}
\tablewidth{0pt}
\tablehead{
\colhead{Objects} &\colhead{Mass\tablenotemark{a}} & \colhead{Momentum\tablenotemark{a}} & \colhead{Kinetic Energy\tablenotemark{a}} \\
\colhead{ } & \colhead{(M$_{\sun}$)} & \colhead{(M$_{\sun}$ km s$^{-1}$)} & \colhead{(10$^{44}$ ergs)}
}
\startdata
Known Outflows  &  10.1 /  51    &  24.7 / 173    &    7.5 / 75  \\
CPOCs                  &  22.3 / 112    &  49.2 / 344    &   13.0 / 130 \\
Total                      &  32.4 / 163    &  73.9 / 517    &   20.5 / 205 \\
\enddata

\tablenotetext{a}{Values before the slash are the original estimates and those after the slash
are corrected values  (see \S~\ref{outmass}).}

\end{deluxetable}

Another way to assess the importance of outflows in driving the turbulence in their local environment is to compare the total outflow energy input rate into the cloud (i.e., outflow luminosity) with the energy rate needed to maintain the turbulence in the gas.  We estimate the outflow luminosity by dividing the outflow kinetic energy by an estimate of the outflow's timescale, $\tau_{flow}$. This term introduces the biggest uncertainty in our estimate of the outflow luminosity as it is hard to determine the age of a molecular outflow solely from molecular line emission data (identification of the outflow source and measurement of the kinematics of the shocks associated with the outflow are needed to obtain a relatively accurate outflow age). 
Most possible candidate
 sources listed in Table~\ref{cpoctab} are protostars in the Class I stage, as reported by Evans et al.~(2009). This stage has an average lifetime of about 0.5 Myr \citep{evans09}, and consequently this provides an upper limit to the typical outflow ages in Perseus. A lower limit on the age of molecular outflows in Perseus is given by the time that a ``typical'' protostellar jet in Perseus has taken to reach its current position.  We estimate this lower limit by assuming a typical jet velocity of 100 km~s$^{-1}$ (Reipurth \& Bally 2001) and a median jet  lobe in Perseus (from infrared Spitzer images) of about 0.3 pc (Guenthner 2009), which results in  a value of $3 \times 10^3$ yr. Here we use a value between these two limits, $\tau_{flow} = 5 \times 10^4$ yr, for our estimation of  the outflow luminosity and caution that there are major uncertainties in this assumption.

 \begin{figure*}
\epsscale{1.1}
\plotone{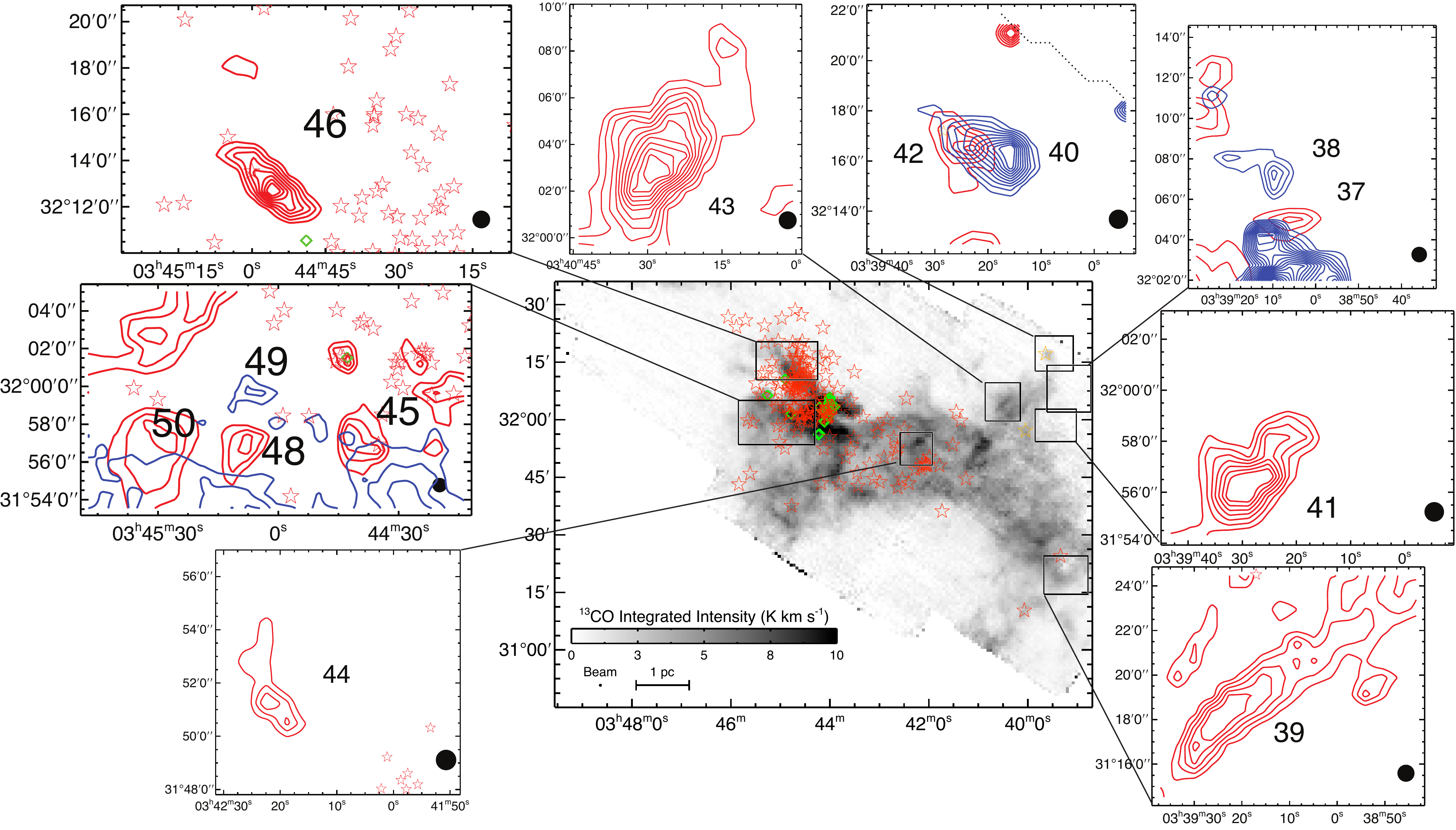}
\caption{CPOCs in Area V. The rest is the same  as in Figure~\ref{fig_reg1}.
\label{fig_reg5}}
\end{figure*}

The numerical study by \citet{maclow99} shows that the energy dissipation of uniformly driven magnetohydrodynamic turbulence is approximately given by:
\begin{equation}
t_{diss} \sim (\frac{3.9\kappa}{M_{rms}})t_{ff}
\end{equation}
where $M_{rms}$ is the Mach number of the turbulence (i.e., the ratio of the turbulence velocity dispersion over the sound speed),
 $t_{ff}=\sqrt{3\pi/32G\rho}$ is the free-fall timescale, and $\kappa = \lambda_d/\lambda_J$, the ratio of the driving wavelength
over the  Jean's length of the clump. Numerical simulations show that the turbulence driving length of a continuous (i.e., non-episodic) outflow is
 approximately equal to the outflow lobe or cavity length (Nakamura \& Li 2007; Cunningham et al.~2009).
In a cluster, different outflows will inject their energy on a range of scales 
(Carroll et al.~2009), and depending on the length and number of episodic events, the driving scale could
vary significantly between outflows. A study of infrared outflows in Perseus using IRAC Spitzer images shows that outflow lobes
in this cloud can range from 0.03 to about 2 pc (Guenthner 2009).
 It is, therefore, difficult to assign a single value to  $\lambda_d$. To simplify our analysis we choose  $\kappa \sim 1$ to 
   obtain an estimate of $t_{diss}$ using the equation above, and warn that for most outflows in our sample this is probably 
   a lower limit as most molecular outflows detected in our maps are at least four beam sizes ($\sim 0.22$~pc) in length and the   
   average $\lambda_J$ of the clumps is $\sim 0.2$ pc.
  
  We estimate the average density of the region using $M_{reg}/R_{reg}^3$, and we assume the temperature of the gas is given by the average cloud excitation temperature in the region (see Table~\ref{physregtab}).  The turbulent dissipation rate (i.e., the power needed to maintain the turbulence in the volume) is then given by $L_{turb} = E_{turb}/t_{diss}$ and it is shown for each region in Table~\ref{physregtab}.  It can be seen that for all six regions the total outflow luminosity is at least 80\% of $L_{turb}$
   and that for five of the regions the outflow luminosity is significantly more ($\geq150\% $) than the power needed
    to maintain the turbulence  (see Table~\ref{comptab}), implying that outflows are an important agent for the maintenance  of turbulence in most regions of active star formation.

    We note that the numerical study from which we obtained equation 3 uses parameters that are appropriate to conditions found in molecular clouds, yet the space dependence of the driving function in these numerical simulations  (which is uniform throughout the simulation cube) differs substantially with the way outflows interact with their parent cloud \citep{maclow99}.  The dissipation rate of the MHD turbulence may vary depending on the space dependence of the driving force, and thus the numbers given in Table~\ref{physregtab} are only rough estimates.  
  In addition, the values of the  
ratio of  total outflow luminosity to energy rate needed to maintain the turbulence in the gas ($r_{L} = L_{flow}/L_{turb}$) are highly uncertain due to the uncertainties in the outflow driving length and timescale. We can constrain the possible range of values of $r_{L}$ by constraining these two uncertain outflow properties. As stated above the value used for $\kappa$ is a lower limit, as the driving lengths of the observed outflows is most likely larger than the assumed length of 0.2~pc. If we assume the sizes of the CPOCs represents the lower limit of the driving length, then  on average the driving length should be at least twice the assumed value of about 0.2~pc.
 With regards to the outflow timescale, it is more probable that $\tau_{flow}$ is less than 0.2 Myr, than it is for it to be closer to the stated absolute upper limit of 0.5 Myr (i.e., the approximate age of the Class I stage). This is because most outflows in Perseus seem to come from Class I sources, and it is very unlikely that all are at the end of that evolutionary stage.
 A larger outflow driving scale, as well as a shorter  $\tau_{flow}$ will increase $r_{L}$.  
 Combing these factors together we can safely deduce that it is very unlikely that the value of $r_L$  is no less that half,
 and could easily be more than,  the value reported  in Table~\ref{comptab}. 
Even with a decrease of a factor of two in $r_L$ we can conclude that outflows in an active region of star formation can be a source of non-negligible power for driving turbulence in the molecular gas, and they should be treated as such in numerical simulations of star-forming clouds.
    
         \begin{figure*}
\epsscale{1.1}
\plotone{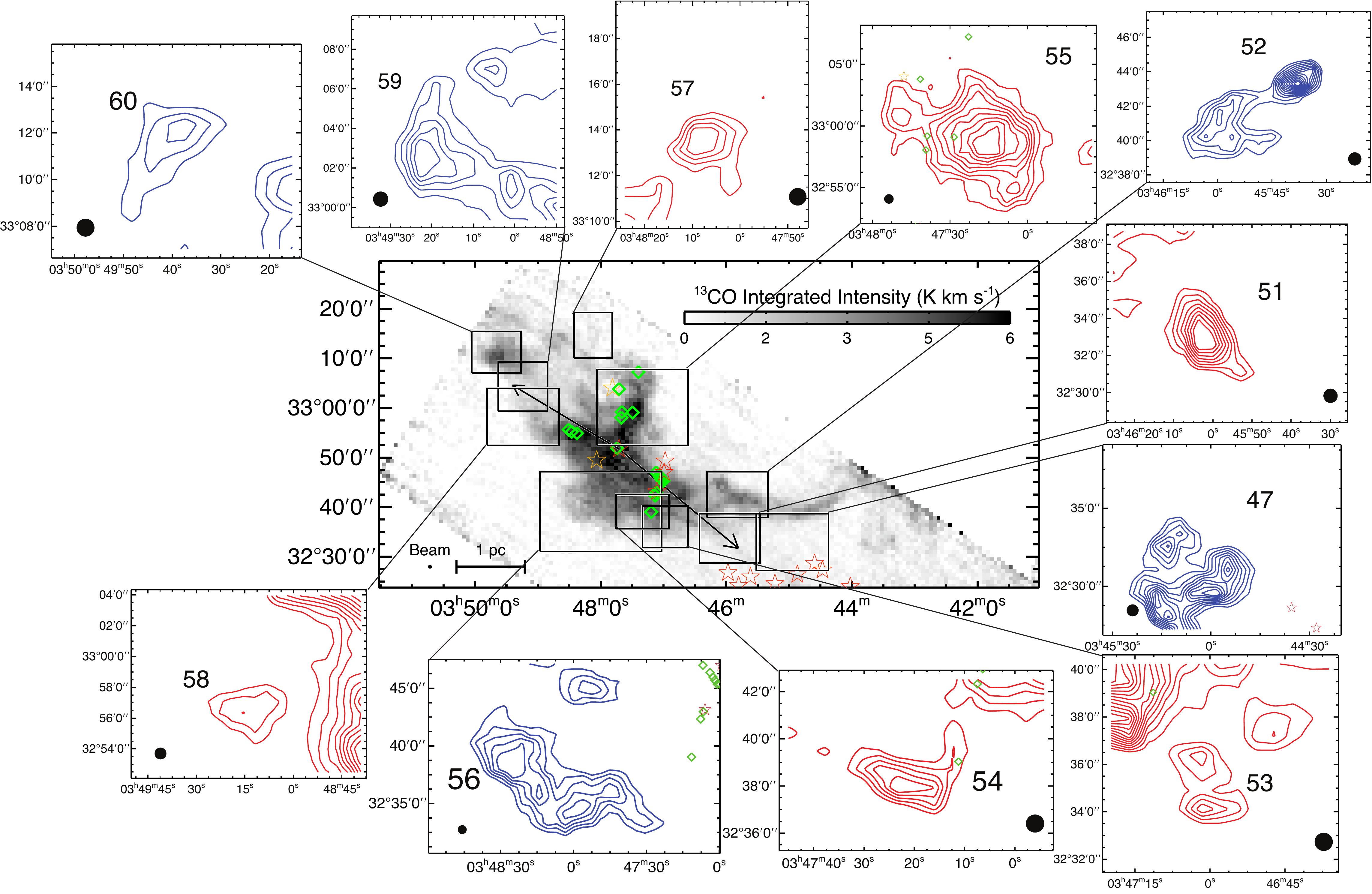}
\caption{CPOCs in Area VI. Lines with arrow show the extent of the B5-IRS molecular outflow lobes, including the new extensions reported here. The rest is the same  as in Figure~\ref{fig_reg1}.
\label{fig_reg6}}
\end{figure*}  

 However, the recent studies of Brunt et al.~(2009) and Padoan et al.~(2009)  have claimed that even in a region full of outflows like NGC~1333  the 
 turbulence is mostly driven by  large-scale mechanisms
 that originate outside the cloud (e.g.,  supornova explosion) and that small-scale driving of turbulence inside molecular clouds may only be somewhat important
 in regions close to outflows. Brunt et al.~and Padoan et al.~use the principal component analysis (PCA) method and the velocity coordinate spectrum (VCS) method, 
respectively, on CO line maps of NGC~1333 to study the turbulent energy spectrum and derive a driving length of turbulence in this region.
 The results of both studies, from  analysis of the COMPLETE $^{13}$CO(1-0) map, imply that  turbulence is not driven at scales smaller than the size of NGC~1333. 
  Padoan et al.~(2009) argue that this suggests the turbulent driving is primarily from sources outside the region rather than outflows within NGC~1333.
   It could very well be that the large-scale shells driven by stellar winds we find throughout the Perseus cloud complex are the source 
   of this large-scale driving mechanism (Arce et al., in preparation). 
  Brunt et al.~(2009)  also analyze the C$^{18}$O(1-0) data of the same region, and they point out that there is evidence for turbulence driving at scales  of  approximately 0.3 to 0.7~pc (assuming a distance to NGC~1333 of 250 pc) which could be due to outflows. We caution that in both of these studies, in order to reach their conclusions, they have to compare the analysis they performed on the observed data to
a similar analysis using  synthesized maps  from numerical simulations. In these simulations turbulence is driven in Fourier space using a field of random Gaussian fluctuations with a limited range of wave numbers. Although this is a reasonable approximation that facilitates running the numerical simulation, it is not how turbulence is driven in nature. Comparing real molecular clouds to simulations that use uniform turbulence driving might provide some information on the driving scale, but it is probably not the whole story. 
As we discussed above, the total outflow energy input rate in the six regions of star formation is enough to drive turbulence. If outflows are not important in driving the observed turbulence, then this raises the question of where does all the outflow energy go. 
It is clear these two studies are not the final word on this issue, and further studies are needed to fully understand the process of outflow-generated turbulence.

\begin{deluxetable}{lcccccc}[h!]
\tablecolumns{7}
\tabletypesize{\scriptsize}
\tablecaption{Regions of Active Star Formation in Perseus
\label{regiontab}}
\tablewidth{0pt}
\tablehead{
\colhead{Name} & \multicolumn{2}{c}{SE corner} &
\multicolumn{2}{c}{NW corner} & \multicolumn{2}{c}{Vel. Range} \\
  &  \colhead{R.A.} & \colhead{Dec}  & 
\colhead{R.A.} & \colhead{Dec}  & \colhead{min.} & \colhead{max.} \\
  &  \multicolumn{2}{c}{(J2000)}  & \multicolumn{2}{c}{(J2000)} & \multicolumn{2}{c}{(km~s$^{-1}$)}
}
\startdata
L1448          & 03:25:42 & 30:37:40 & 03:23:55 & 30:53:00 & 3.3 & 6.3\\     
NGC1333 & 03:31:05 & 30:33:50 & 03:26:52 & 31:59:40 & 5.2 & 10.2\\
B1-Ridge      & 03:33:07 & 30:30:50 & 03:30:19 & 30:53:40 & 4.6 & 8.5\\
B1                 & 03:34:00 & 30:54:30 & 03:31:33 & 31:19:50 & 4.8 & 8.7\\
IC348         & 03:46:02 & 31:52:50 & 03:43:06 & 32:29:40 & 6.5 & 11.4\\ 
B5                  & 03:51:05 & 32:30:20 & 03:44:07 & 33:28:40 & 8.7 & 11.4\\
\enddata

\end{deluxetable}

\subsubsection{Outflows and Star Formation Efficiency}
\label{outsfe}

Recent  three-dimentional numerical simulations  show that outflows from a cluster of protostars can easily sustain the supersonic turbulence in the surrounding gas
(Nakamura \& Li 2007, herafter NL07; Carroll et al.~2009). In particular, the results of NL07 show that the turbulence in an isolated cluster-forming region of about two parsecs in size, fed by the energy and momentum injected by bipolar outflows, can maintain the region close to dynamic equilibrium. In this quasi-equlibrium state,  infall and outflows motions are approximately equal, but there is a net flux of mass towards the bottom of the potential well. This results in a slower collapse, compared to the global free-fall time, and only  a few percent of the total gas mass is converted into stars within a free-fall time. Different runs by NL07 with differing outflow strengths show that the star formation rate per free-fall time decreases with increasing average outflow strength. Their study shows that more powerful outflows result in an increase in the turbulence in the gas, which then leads to a delay of the gravitational collapse and, consequently,  
a lower star formation efficiency per free-fall time. 

\begin{deluxetable*}{lccccccccc}
\tablecolumns{10}
\tabletypesize{\small}
\tablecaption{Physical Parameters of Active Star Forming Regions in Persesus
\label{physregtab}}
\tablewidth{0pt}
\tablehead{
\colhead{Name} & \colhead{$M_{reg}$\tablenotemark{a}} &  \colhead{$R_{reg}$\tablenotemark{b}} &
\colhead{$\Delta v$\tablenotemark{c}} & \colhead{$T_{ex}$\tablenotemark{d}}   
& \colhead{$v_{esc}$\tablenotemark{e}} & \colhead{$E_{grav}$\tablenotemark{f}} 
& \colhead{$E_{turb}$\tablenotemark{g}} &\colhead{$t_{diss}$\tablenotemark{h}} & \colhead{$L_{turb}$\tablenotemark{i}}\\
\colhead{} & \colhead{(M$_{\sun}$)} &  \colhead{(pc)}  & \colhead{(km~s$^{-1}$)}  &  \colhead{(K)} &
  \colhead{(km~s$^{-1}$)} & \colhead{($10^{46}$~erg)}
& \colhead{($10^{45}$~erg)} & \colhead{($10^5$ yr)} &  \colhead{($10^{32}$~erg~s$^{-1}$)}
}
\startdata

L1448            &  150 & 0.6 & 1.9 & 10 & 1.5 &  0.3  & 2.9  &  2.6 &  3.6\\
NGC~1333    & 1100 & 2.0 & 2.2 & 13  &  2.2  & 5.2  & 28.8   &  5.7 & 15.9\\
B1-Ridge      &  210 & 0.7 & 1.9 & 13 & 1.6  &  0.5  & 4.1 &   3.1 & 4.1\\
B1                  &  430 & 0.9  & 2.1 & 13  & 2.0  & 1.8  & 10.2 &   2.9 & 11.2\\
B5                  &  420 & 1.4 & 1.5  & 12 & 1.6 & 1.1   & 5.1   &   7.6 & 2.1\\
IC~348            &  620 & 0.9 & 1.8 & 15  &  2.4  &  3.7  & 10.9  &  3.0 &   11.4\\

\enddata

\tablenotetext{a}{Mass of star-forming region, obtained using the procedure described in \S~\ref{outmass}}
\tablenotetext{b}{Radius estimate of the region obtained from the geometric mean of minor and major axis
of the extent of the $^{13}$CO integrated intensity emission.  }
\tablenotetext{c}{Average velocity width (FWHM) of the $^{13}$CO(1-0) line in the region.}
\tablenotetext{d}{Average excitation temperature of region.}
\tablenotetext{e}{Escape velocity, given by $\sqrt{2GM_{reg}/R_{reg}}$.}
\tablenotetext{f}{Gravitational binding energy given by $GM_{reg}^2/R_{reg}$.}
\tablenotetext{g}{Turbulence energy given by $\frac{3}{16~ln2}M_{reg}\Delta v^2$.}
\tablenotetext{h}{Turbulence dissipation time,  see~\S~\ref{outturb}}
\tablenotetext{i}{Turbulence energy dissipation rate give by $E_{turb}/\tau_{diss}$.}
\end{deluxetable*}

Here we investigate whether there is any relationship between the star formation efficiency and the total outflow strength in different star-forming regions in the Perseus cloud. We use the c2d catalog of young stellar objects (Evans et al. 2009) to search for the total number of young stars in each of the six different star-forming regions in Perseus studied here (see Table~\ref{regiontab} and Figure~\ref{perseus_map}). The total number of YSOs within the area of each region, as well as the fraction of those that are embedded (i.e., Class 0 and I) sources, are shown in Table~\ref{sfetab}. For each region, 
we calculate the current star formation efficiency ($SFE$)\footnote{Note other studies mostly use $SFE = M_{YSO}/(M_{reg} + M_{YSO})$ 
(e.g., J\o rgensen et al.~2008; Evans et al.~2009). This formula does not take into consideration the gas that used to be in the cloud but that has been lost due to different processes
triggered by the star formation process, other than the mass that is used to directly form the star.} 
using  $SFE = M_{YSO}/(M_{reg} + M_{YSO}+f*M_{flow})$, where $M_{YSO}$ is the total mass of young stellar objects within the region's boundaries. We assume an average YSO mass of 0.5 M$_{\sun}$, as in
J\o rgensen et al.~(2008) and Evans et al.~(2009). 
The mass of the gas in the region is given by
$M_{reg}$  (see Table~\ref{physregtab}), and  $f*M_{flow}$ is the mass ejected (or lost) from the region by outflows and other mechanisms. The observed total outflowing mass is given  by $M_{flow}$ (see Table~\ref{outregtab}) and $f$ is a correction factor accounting for mass no longer visible (e.g., from previous outflow events, due to dissociation, etc.). It is difficult to obtain a precise estimate of $f$, but we think that a reasonable estimate lies in the range of 2 to 10, depending on the age of the star forming region (see below).

A cloud with a constant star formation rate will exhibit a higher star formation efficiency as the cloud evolves, since more of the
cloud gas will be transformed into stars. 
 In fact, the results of NL07 show that independent of average outflow strength, as soon as the first stars are formed, the SFE increases 
approximately linearly with time. In addition, regions with different free-fall times (and the same turbulent energy) will collapse at different rates and thus will exhibit different values of the current SFE (even for clouds of the same age).
We therefore need to take into consideration the length of time that the region has been forming stars and its free-fall time in order to investigate whether the star formation efficiency of  different regions depends on the outflow strength.\footnote{The magnetic field strength will also have an effect on the star formation efficiency.  NL07 find a clear trend where SFE decreases with increasing magnetic field strength. Hence, in principle one should also take into consideration the magnetic field in order to compare the SFE of different clouds. 
Unfortunately only two of the regions studied here (B1 and L1448) have confirmed detection of the {\it line-of-sight} component of the magnetic field 
(Goodman et al.~1989; Troland \& Crutcher 2008). For simplicity (and lack of measurement of the magnetic field strength in all the Perseus star-forming regions), we will assume that all regions have the same magnetic field strength.}
For this purpose, we define the ``normalized'' star formation efficiency, $SFE_n \equiv SFE*(t_{ff}/\tau_{SF})$, where 
$t_{ff}$ is the free-fall timescale  of the region (see above and Table~\ref{sfetab}), and $\tau_{SF}$ is the time the region has been forming stars (i.e., the region's age). 

\begin{deluxetable}{lcccc}[!b]
\tablecolumns{5}
\tabletypesize{\small}
\tablecaption{Total Outflow Mass, Momentum, Energy and Luminosity in Star-Forming Regions
\label{outregtab}}
\tablewidth{0pt}
\tablehead{
\colhead{Name} & \colhead{$M_{flow}$\tablenotemark{a}} & \colhead{$P_{flow}$\tablenotemark{a}} & 
\colhead{$E_{flow}$\tablenotemark{a}} & {$L_{flow}$\tablenotemark{b}}\\
\colhead{}  &  \colhead{(M$_{\sun}$)} & \colhead{(M$_{\sun}$~km~s$^{-1}$)} 
  & \colhead{($10^{44}$~erg)}  & \colhead{($10^{32}$~erg~s$^{-1}$)}
}
\startdata
L1448           &  1.0 / 5       &          3.1 /  21.7      &      1.2 / 12      &             8 \\
NGC1333    &  5.0 / 25    &        17.4 / 121.8    &      6.9 / 69      &            44 \\  
B1-Ridge     &  1.1 / 5.5   &          3.2 /  22.4      &     1.0 / 10       &              6 \\ 
B1                 &  1.5 / 7.5    &         6.2 /  43.4      &    3.1 / 31         &            20 \\
IC348          &    4.2 / 21    &         7.7 /  53.9      &    1.5 / 15          &           10  \\ 
B5               &    12.8 / 64    &         22.3 / 156.1   &     4.1 / 41        &           26  \\ 
\enddata
\tablenotetext{a}{Values before and after the slash are the original estimates and the
estimates adjusted by  the correction factor, respectively (see \S~\ref{outmass}).}
\tablenotetext{b}{Outflow luminosity, $L_{flow} = E_{flow}/\tau_{flow}$, 
obtained using the value of the total outflow kinetic energy adjusted by the correction 
factor and using an average outflow timescale of $5\times10^4$~yr}
\end{deluxetable}

  \begin{deluxetable*}{lccccc}[b!]
\tablecolumns{6}
\tabletypesize{\small}
\tablecaption{Quantitative Assessment of Outflow Impact on  Star-Forming Regions
\label{comptab}}
\tablewidth{0pt}
\tablehead{
\colhead{Name} & \colhead{$E_{flow}/E_{turb}$} & \colhead{$r_L = L_{flow}/L_{turb}$} & 
\colhead{$E_{flow}/E_{grav}$} &  \colhead{$M_{esc}\tablenotemark{a}$ [M$_{\sun}$]} & \colhead{$M_{esc}/M_c$} 
}
\startdata

L1448              &     0.41     &     2.1    &    0.40    &    15   &  0.10 \\
NGC1333       &     0.30     &     3.4    &     0.17    &    76   &   0.07 \\
B1-Ridge      &       0.24     &     1.5    &     0.20    &    14   &   0.07 \\
B1                  &       0.30      &     1.7     &     0.17   &      21  &   0.05  \\
IC348            &       0.14     &     0.8     &     0.04   &     23  &   0.04 \\
B5                 &        0.80     &    12.4    &     0.37   &     98   &   0.23 \\

\enddata
\tablenotetext{a}{Escape mass, given by $M_{esc} = P_{out}/v_{esc}$ (see \S~\ref{outdisp}).}
\end{deluxetable*}

Studies of the population of young stars in IC~348 (Luhman et al. 2003; Muench et al.~2007) and NGC1333 (Lada et al.~1996; Wilking et al.~2004) indicate that these two clusters have significantly different ages of about 2-3 Myr and  about 1 Myr, respectively.  Here we will assume an age of 2.5 Myr for IC~348 and 1 Myr for NGC~1333.
For the other clumps in Perseus we use the fraction of Class I sources in the region to estimate their age. Assuming that star formation has been approximately constant, then the fraction of embedded protostars in a region can provide a rough relative lifetime of the star formation process in the clump. In IC~348 and NGC~1333 the fraction of Class I sources is 10\% and 32\%, respectively (see Table~\ref{sfetab}), consistent with the age difference between these two clusters (J{\o}rgensen et al.~2006). Similar to NGC~1333, all the other star-forming regions in Perseus, except L1448, have a fraction of Class I sources of about 30\% to 50\% and their YSO populations include evolved young stars (with ages of about 1 Myr) in the late Class II stage ---as derived from the sources' bolometric temperature and spectral index (see Evans et al. 2009 for more detail).  We thus assume that regions B5, B1, and B1-Ridge also have an age of about 1 Myr. In L1448  all the YSOs are Class I sources, which suggests that this region is significantly younger than the rest. We assume the age of L1448 is 0.5 Myr, the average lifetime of the Class I stage as derived by Evans et al.~(2009). 
 We also use the relative populations of embedded protostars to obtain a rough estimate of  $f$, the correction factor accounting for mass no longer visible in the $SFE$ equation above. The value of $f$ in L1448 should be between 1 and 2, 
 while $f\sim3$ should be a reasonable estimate for
  B5, B1, B1-Ridge and NGC~1333. In IC~348 star formation  has been taking place for a longer time ,and thus use $f=8$ for this region.

\begin{figure}[!h]
\epsscale{1.3}
\plotone{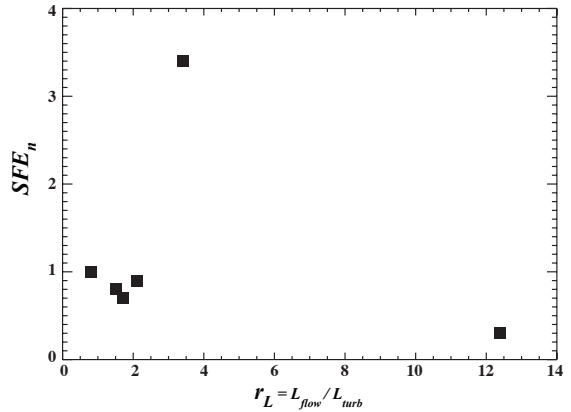}
\vspace{-1.35in}
\caption{Normalized star formation efficiency as a function of $r_L$ for different regions of star formation in Perseus. No error bars are shown as it is hard to estimate accurate $1-\sigma$ uncertainties for these points. The uncertainties in the values of $SFE_n$ and $r_L$ could easily be a factor of two (or even more).
\label{sfefig}}

\end{figure}

The values of the normalized SFE in the different star-forming regions of Perseus, shown in Table~\ref{sfetab}, range from 0.3\% to 3.4\%, with B5 and NGC1333 showing the smallest and largest values, respectively.   From the study of NL07 we expect a lower $SFE_n$ for regions with stronger outflows. 
We use  the 
ratio of the total outflow luminosity to the energy rate needed to maintain the turbulence in the gas ($r_{L}$, in Table~\ref{comptab}) as a way to quantify the outflow strength and their impact  on the cloud's turbulence.  
Clouds with $r_{L} > 1$ harbor outflows that input enough energy into the surrounding gas to maintain the observed turbulence in the cloud. 
Assuming the magnetic field strength is approximately the same for all clouds in Perseus, then according to NL07, the outflow-driven turbulence should be the dominant source for variations in the SFE.  We would therefore expect clouds with a high  $r_{L}$  to have a low $SFE_n$.  However, our results do not show a clear dependence of $SFE_n$ with $r_L$ (see Figure~\ref{sfefig}).
We see that B5 (the region with the largest $r_{L}$)  exhibits the lowest value of $SFE_n$ but otherwise there is no correlation between $r_{L}$ and $SFE_n$.  
 
 
   It is hard to draw any  strong conclusion from our results probably due to the large uncertainties in estimating $SFE_n$. The number of stars formed in the cloud (a number essential for estimating SFE) is uncertain as we only include YSOs in the c2d catalog that are within
  the regions defined by us, and thus we are not accounting for stars formed in the cloud but 
  that have since moved away from the defined region.
   In addition, the list of young stars is not 100\% complete as, among other things, it does not include pre-main sequence stars that do not have a detectable infrared excess (see Evans et al.~2009 more details). Moreover, age estimates for the star-forming regions, as well as the free-fall timescales are probably uncertain within a factor of two or so. 
 Given the large uncertainties in $SFE_n$, the small size of the sample and the limited range in values of the total outflow momentum and $r_L$ among all regions, it is of no surprise that we do not find a significant correlation between $SFE_n$ and 
outflow strength, even if such correlation exists.  Clearly, a larger sample of star-forming regions with accurate age and SFE estimates is needed in order to study this further.


\subsubsection{Disruption of the cloud by outflows?}
\label{outdisp}
It has also been proposed that outflows can have a disruptive impact on their clouds \citep[e.g.,][]{arce02a}. 
One way to estimate the disruptive effects of outflows on clouds is by comparing the outflow energy with the cloud's gravitational energy.  From Table~\ref{comptab} it can be seen that the total kinetic energy of the outflows is only a small fraction (between 4 and 40\%) of the gravitational binding energy of each region.  From this simple analysis, it appears that for most regions, outflows do not have the necessary energy to significantly disrupt their cloud.  Another way to assess the potential disruptive effect of outflows on clouds is using the ``escape mass'' ($M_{esc}$) of the region, defined as the mass that could potentially be dispersed by the outflows if we assume 
that the total current outflow momentum is used to  accelerate a total gas mass of
$M_{esc}$, through conservation of momentum, to the cloud sub-region's  escape velocity (i.e.,   $M_{esc}=P_{flow}/v_{esc}$).  We note that in all regions the {\it current} total outflow mass is lower than $M_{esc}$ only by factors of 1.5 to 3.5, so it is possible that the current outflows could eventually entrain as much mass as $M_{esc}$. From our analysis we see that outflows in the active star-forming regions of Perseus currently have enough momentum to potentially disperse only 4\% to 23\% of the mass in their respective regions (see Table~\ref{comptab}).

\begin{deluxetable}{lcccccc}[b!]
\tablecolumns{7}
\tabletypesize{\scriptsize}
\tablecaption{Star Formation Efficiency in Different Regions of the Perseus Molecular Cloud
\label{sfetab}}
\tablewidth{0pt}
\tablehead{
\colhead{Region} & \colhead{No.} & \colhead{No.} & \colhead{SFE\tablenotemark{b}} & \colhead{$t_{ff}$\tablenotemark{c}} & 
\colhead{$\tau_{SF}$\tablenotemark{d}} & \colhead{$SFE_n$\tablenotemark{e}}\\
\colhead{Name} & \colhead{YSOs} & \colhead{Embedded\tablenotemark{a}} & \colhead{(\%)} & \colhead{Myr} & \colhead{Myr} & \colhead{(\%)}
}
\startdata

L1448               &   5                                 &  5  (100\%)    &  1.5   &   0.31  &  0.5   &  0.9\\
NGC1333      & 125                              & 40   (32\%)    &  4.9   &   0.69  &  1     &  3.4\\
B1-Ridge         &  11                               &  3   (27\%)    &  2.4   &   0.33  &  1     &  0.8\\
B1                      &  19                              & 10   (52\%)    &  2.1   &   0.33  &  1     &  0.7\\
IC348             & 154                             & 16   (10\%)    &  9.0   &   0.28  &  2.5   &  1.0\\
B5                     &   5 \tablenotemark{f} &  3   (60\%)    &  0.4   &   0.65  &  1     & 0.3\\

\enddata
\tablenotetext{a}{Number of Class 0 and Class I sources (using $T_{bol}$ classification scheme, see Evans et al.~2009) and percentage of total YSOs in parenthesis.}
\tablenotetext{b}{Star formation efficiency, $SFE=M_{YSO}/(M_{reg}+M_{YSO}+f*M_{flow})$, see \S~\ref{outsfe}}
\tablenotetext{c}{Free-fall timescale}
\tablenotetext{d}{Approximate time region has been forming stars}
\tablenotetext{e}{$SFE_n = SFE * t_{ff}/\tau_{SF}$}
\tablenotetext{f}{The c2d catalog gives four candidate YSOs in this area, but the {\it Spitzer} observations do not cover the entire B5 region. Only one previously known YSO lies in the
unmapped area. We added this source for a total of 5 YSOs in B5. This is a lower limit as a few low-luminosity YSOs associated to B5 could lie in the area that was not
covered by the c2d survey.}
\end{deluxetable}

It appears that current outflows do not have enough strength 
 to cause serious impact to the integrity of their cloud. However, 
it is possible that the estimated momentum from our observations of the current outflows is just a fraction 
 of the total momentum from all outflow events from
 all protostars that form throughout the entire life of the cloud. 
   As discussed above, the observed current molecular  outflows most probably have ages of about 0.2 Myr or less.
 In addition,  most of the star forming
  regions in Perseus are only $\sim1$ to 3 Myr old, but molecular clouds have lifetimes of about 3 to 6 Myr (e.g., Evans et al.~2009). 
If star formation and outflow production continues at a roughly constant rate (or increases with time), it is then
possible that the total momentum injected by outflows into the gas, throughout the entire history of the star-forming region 
could be considerably larger that  
what we estimate from our limited snapshot of the cloud.  
If, for example, the total outflow momentum injected over the lifetime of the cloud turns out to be 
a factor of ten more than the current observed outflow momentum,
this would then imply that outflows could seriously disrupt the three least massive regions in our study 
(i.e, L1448, B1-Rdige, and B5), as outflows could potentially disperse 70\% or more of the current mass in these regions.
In this scenario, outflows in the other three region would be able to potentially dispersed 40 to 70\% of the gas. 
A loss of 40\% of the original mass may still cause some significant damage to the integrity of the clump as it will be left 
 with a gravitational binding energy that is 36\% of the original value (assuming the cloud size stays the same).
 We conclude that outflows may disperse some gas from their parent cloud, yet
  unless the amount of momentum that outflows inject into their surroundings throughout the entire lifetime of the cloud
  is more than a factor of ten compared to current outflow momentum estimates, 
  outflows cannot fully disrupt their parent clouds.

The theoretical work of Matzner \& McKee~(2000) studied the combined 
effects of many outflows on the gas of clumps forming clusters of low-mass stars. 
Their results predict about  30\% to 50\% of the mass will be turned into stars, while the rest 
(i.e.,  50\% to 70\%) of the gas in the cluster-forming region will be ejected by 
 outflows. For these theoretical results to be consistent with our findings, 
 outflow momentum injection throughout the entire  lifetime of the cloud would have to be more than 
 a factor of 10 of the current outflow momentum for all 
 studied regions in Perseus to experience (at least) a 50\% gas dispersion due to  outflows by the end of cloud's lives.
 It seems very unlikely that in all cases outflows alone will disperse such a high fraction of the cluster-forming
 gas. A more plausible scenario would be that outflows help disperse a fraction of their surrounding gas
 and other mechanisms, such as dispersion by stellar winds and erosion produced by  radiation,  
 help dissipate the rest of the gas that does not end up forming stars ---thereby producing the
  star formation efficiencies ($\sim 10\%$ to 30\%) 
  observed in clusters (e.g., Lada \& Lada 2003).

 We note that the discussion above only relates to the impact of outflows on their cloud at scales of about 1~pc, and it does
 not pertain to the impact of outflows on their host cores.  
 Our observations
probe regions with size of $\sim 1$ to 4 pc and density of $\sim 10^3$, which are 
 are considerably more extended and less dense than the  cores in Perseus,
 which have an average radius and density of  $\sim 0.04$~pc and $\sim 10^5$ cm$^{-3}$ (Enoch et al.~2008). 
The results presented here {\it do not} rule out 
that outflows can be a major player in the 
mass loss process in cores (as reported by, e.g.,  Fuller \& Ladd 2002; Arce \& Sargent 2005, 2006).

\section{Summary and Conclusions}

A survey of all the high velocity outflow features was conducted for Perseus using $^{12}$CO and $^{13}$CO molecular line maps from the COMPLETE Survey in order to gain a better understanding of how outflows effect their host cloud complex over large areas. Our $\sim 6.25\arcdeg \times 3\arcdeg$  maps cover almost the entire Perseus molecular cloud complex which allows us to search for outflows throughout the extent of the cloud and not just in the immediate surrounding of active star-forming regions as in most previous studies.
 We utilize a novel way to search for outflows using 3D visualization of the molecular cloud in position-position-velocity space. Rendering the molecular line data this way  shows high-velocity blobs as clearly visible spikes that stick out from the general cloud gas distribution. Using this technique we detected previously known outflows and new outflow candidates.
 Our survey results show that most molecular outflows 
in Perseus concentrate close to a few groups and clusters of protostars,  and there are
vast areas of the cloud where there is little or no molecular outflow activity.
 We compiled a list of 60 COMPLETE Perseus Outflow Candidates (CPOCs) that lie 
across the entirety of Perseus, with most of them found around the periphery of, and in between, active star forming regions.
We identified a few of these CPOCs as being previously unknown extensions of  previously known outflows. 
In particular, we find new extensions to the well-known  B5-IRS1
outflow which now has a total projected length of $\sim 1\arcdeg$ (or about 4.4 pc).   

The newly identified candidate outflows more than double the amount of molecular outflow mass, momentum and kinetic energy in Perseus. Outflows have considerable impact on the environment immediately surrounding 
 localized regions of active star formation, but 
 lack the energy needed to feed the observed turbulence in the entire Perseus cloud complex, even with the increase 
 in outflowing material found by our study.
We propose there must be  
another energy source, in addition to collimated outflows from protostar, responsible for maintaining the turbulence on a global cloud scale. We leave the
discussion on spherical winds from young stars as a possible source of turbulence in Perseus for a subsequent paper.

We compared the total outflow momentum, energy  and power in the different local star-forming regions 
---areas of active star formation with numerous outflows and sizes 1 to 4 pc---  with the
 energetics of the host cloud sub-region to quantitatively assess the outflow
impact on their surroundings. Our results indicate that outflows
 have enough power (or can contribute substantially) to maintain the turbulence in their local environment.
  
 The numerical simulations  of Nakamura \& Li (2007) show that outflow-driven turbulence can affect the star formation efficiency in their parent cluster-forming clump and we use 
our data (together with data from the literature) to investigate if this is the case in Perseus. 
We detect no correlation between star formation efficiency and outflow strength, but the large uncertainties in these values prevent us from reaching any definitive conclusion on this issue.

Our quantitative assessment of the impact of current  outflows on Perseus indicates that outflows have the potential to disperse and unbind some mass from the cloud.  
The current outflow energy and the amount of material that current outflows can potentially unbind, compared to the total mass of the region, is not enough to cause major disruptions to the cloud. The total outflow momentum injection throughout the entire lifetime of the cloud would have to be more than an order of magnitude greater than the 
 detected current outflow momentum in order for gas dispersion by outflows to cause a serious impact to the integrity of their host cloud.
We argue that, in addition to collimated outflows from young stars, other mechanisms of cloud dispersal are needed to explain the low star formation efficiency in clusters.

 
 


\acknowledgments

We would like to thank to Mark Heyer for his help in the reduction of the CO data, and the anonymous referee for very useful comments that helped improve the quality of the paper. 
We also would like to thank the entire COMPLETE Team for their help with the acquisition and reduction of the data, 
 as well as Adam Frank, Eric Blackman, and Jonathan Carroll for very useful discussions on outflow-driven turbulence.
 Our gratitude goes to the Initiative in Innovative Computing at Harvard for their support of the Astronomical Medicine Project. H.~G.~A. was partially funded by NSF awards AST-0401568 and AST-0845619 while conducting this study. The COMPLETE Survey of Star Forming Regions is supported by the NSF grant No.~AST-0407172.  Research and development of 3D~Slicer is funded by the National Alliance for Medical Imaging Computing NIH Roadmap Initiative grant U54~EB005149, and the Neuroimage Analysis Center grant P41-RR13218.

{\it Facilities:} \facility{FCRAO}.

\appendix
\section{Description of CPOCs}
Here we discuss our findings for each of the six areas we divided the Perseus cloud complex into (as shown in Figure~\ref{perseus_map}).

\subsection{Area I}
Area I is the westernmost part of the Perseus cloud complex, which includes the L1448 and L1455 dark clouds. L1448 is a relatively small ($\sim$0.5 deg$^{2}$, 150 M$_{\sun}$) and isolated star forming region centered around $\alpha = 03^h25^m30^s$, $\delta = 31\arcdeg45\arcmin$. This region  harbors five embedded (Class 0 and Class I) protostars, and no Class II or III sources \citep{evans09}.  The protostars in L1448 power well-known outflows that have been observed across multiple wavelengths \citep{bachiller90L1448,bal97,wolfchase00,eis00,wal05, kwon06, tobin07,davis08}. 
The L1455 region 
lies about one degree southeast of L1448,  and it is only partially included in the lower edge of our CO maps. 
L1455 harbors protostars that power previously-observed molecular outflows, but with our limited map of the region we only detect 
 the outflow that has a southeast-northwest axis and is associated with L1455-IRS4 \citep{bal97,jor06}.

We detected four new candidate outflow features in the periphery of the L1448 region (CPOCs 1 to 4) and one in the area between L1448 and L1455 (CPOC 5).  
These high-veolcity features were marked as candidate outflows because they show ``typical'' outflow spectra with low-intensity high-velocity wings and they lie in close proximity to an active region of star formation with multiple HH objects in their surroundings.  
 CPOCs 1, 2 and 3 lie west-northwest of the center of the L1448 cluster of protostars and lie close to several HH objects  in this region (i.e., HH194 and HH268). The morphology and position of CPOCs 1 and 2 (see Figure~\ref{fig_reg1}) strongly suggest that they are extensions  
 of the outflow presumably powered by IRAS 03220+3035, also known as L1448-IRS1 \citep{bal97}. CPOC 3 is redshifted and lies about 10\arcmin \/ north of
 L1448, and within 3\arcmin \/ of two groups of HH objects.
 CPOC 4 is blueshifted and lies less than 10\arcmin \/ southwest of L1448-IRS1. 
 The position and velocity of this CPOC are not consistent with it  being powered by L1448-IRS1,
  but given its close proximity to the group of embedded  protostars in L1448, is highly probable that this CPOC is powered by any one of these forming stars. CPOC 5 is 
blueshifted and lies just south of the c2d YSO candidate SSTc2dJ032519.52+303424.2. It is not clear whether this YSO candidate or any of the protostars in L1448 or in L1455 is the powering source of this CPOC.

\subsection{Area II}
This area mostly includes the NGC~1333 cluster and its surroundings. NGC~1333 is presently the most active star forming region in Perseus, covering $\sim$1 deg$^{2}$ with a total gas mass of 1100 M$_{\sun}$.  Located at the western part of the cloud complex, it contains two contiguous  young stellar clusters that include about  275 members ranging in age from protostars to pre-main-sequence stars (i.e., Class 0 to III sources) \citep{lada96,reb07,guter08}.  NGC~1333 has also been surveyed for outflows at different wavelengths \citep{bal96a,yan98,knee00,davis08,hat09}.  This region contains several identified and well-studied molecular outflows including that of IRAS 4A, IRAS 2A, IRAS 2B and HH7-11   \citep{bach00,knee00,jor04,choi06, jor07}.  We detect high-velocity CO emission at positions coincident with these outflows, however due to the relatively low 
resolution of our map  we cannot disentangle individual outflows in the central core of the 
NGC~1333 cluster ---the area within $\alpha$ from about $03^h28^m30^s$ to $03^h29^m30^s$ and within $\delta$ from about
$31\arcdeg10\arcmin$ to $31\arcdeg30\arcmin$ (J2000), studied previously by Knee \& Sandell (2000) and Hatchell \& Dunham (2009).

Although NGC~1333 has been very well surveyed for molecular outflows in its central clustered region, few studies have examined the cluster's outskirts.  We identify new candidate outflows  all around the outer parts of NGC~1333 as well as in the gas east of the cluster, between NGC~1333 and B1, and near the Per~6 protostellar aggregate south of NGC~1333 \citep{reb07}. In the western edge of the cluster  lies CPOC 6, a high-velocity redshifted candidate outflow close to HH~338 (see Figure~\ref{fig_reg2}). CPOC 7 is a blueshifted outflow candidate south of the main cluster core, and close to several HH objects and YSO candidates (the closest being SSTc2dJ032834.49+310051.1). 
CPOC 8, located south of NGC~1333,  lies 
 between three c2d YSO candidates and two HH objects (HH 750 and HH 743), but it is not clear which of these nearby young stars is the source of this elongated outflow candidate. 
 In the west-southwest outskirts of the NGC~1333 core lies CPOC 9, a redshifted outflow candidate that coincides with the position of at least five HH objects 
 (and H$_2$ knots) and two YSO candidates (SSTc2dJ032832.56+311105.1, and SSTc2dJ032837.09+311330.8). 
 CPOCs 13 and 10 lie to the immediate north and northwest of the cluster core, respectively. The most likely powering source of CPOC 10 is SSTc2dJ032844.09+312052.7, which lies on the southeast corner of this outflow candidate. CPOC 13 is coincident with  at least six HH objects and H$_2$ knots
 and one of the multiple sources just south of this CPOC could be its powering source. 
  CPOCs 11 and 12 are  redshifted and blueshifted, respectively, and lie very close to each other. There are no known YSOs right at the position between these two CPOCs, so it is likely that each outflow candidate is associated with a different powering source. From the distribution of the high-velocity gas in relation to the position of the various YSO candidates in the region,  STTc2dJ032834.53+310705.5 seems 
  to be the most likely candidate for being the powering source of CPOC 11, while SSTc2dJ032843.24+311042.7 is the most
  likely candidate source of CPOC 12.
 CPOCs 14 and 15 are another pair of  red and blue outflow candidates that lie near HH 757A/B and HH 426 as well as near two c2d YSO candidates in the southern outskirts of the NGC~1333 cluster. 
In this region we also detect high-velocity redshifted emission located at the same position as IRAS~03254+3050, just south of the main cluster core. 
We do not include this in the list of CPOCs as this high-velocity gas is coincident with the high-velocity CO(3-2) emission recently reported by \citet{hat09}. We note that the prevalence of redshifted CPOCs north and west of NGC~1333 is due to the fact that there is intervening gas at blueshifted velocities with respect to the central velocity of NGC~1333 that hampers our ability to detect outflow-related blueshifted emission. 
  
In the southern edges of the cluster core, and coincident with HH 18A and  H$_2$ knot 25 in Davis et al.~(2008), we find the redshifted CPOC 16. It has a northwest-southeast elongation which indicates it could be part of an outflow with its origin near the cluster center. CPOC 17 is a high-velocity redshifted blob east of the cluster core. Its relatively high mass and size implies that it is probably the result of the impact of multiple outflows on the surrounding molecular gas. This picture is supported by the fact that there are multiple YSO candidates and HH objects (including the HH flows HH 497 and HH 336) coincident with CPOC 17. 
On the northeastern outer limits of the main NGC~1333 cluster, and coincident with HH~764, lies CPOC 18. About five arcminutes  east  of  CPOC 18, and close to HH~497,  lies CPOC 20.
It is very probable that this CPOC is the extension of one of the many molecular outflows originating in NGC~1333.  
To the north of NGC~1333 we also find CPOC 19, which has a highly elongated morphology, typical of collimated outflow lobes, and lies  close to SSTc2dJ032923.48+313329.5 (IRAS~03262+3123).  
CPOC 21, in the eastern outskirts of the main part of the NGC~1333 cluster, has a west-east elongation and it is coincident with HH~767. This HH object is part of a chain of HH objects that presumably are part of the same east-west parsec-scale flow, composed of HH~348, HH~349, HH~766, and HH~767 (Walawender et al.~2005). 
A candidate powering source for this CPOC is SSTc2dJ033024.08+311404.4, which lies at the eastern edge of the redshifted gas.
Halfway between the B1 and NGC~1333 star forming regions we find CPOCs 23, 25 and 27. 
 It is hard to tell whether this chain of high-velocity blueshifted blobs are related (or are driven by different sources) and whether they are part of an outflow originating in NGC~1333 or B1. Lastly, in this region we find CPOC 22, a blueshifted candidate outflow that most likely is driven by one of the protostars in the recently discovered Per~6 aggregate \citep{reb07}.

\subsection{Areas III and IV} 
This part of the Perseus cloud complex includes the 
Barnard 1 (B1) dark cloud, the ridge southwest of B1 (the B1-Ridge), and the area just east of  B1 (see Figure~\ref{perseus_map}).   B1 and the B1-Ridge are two  regions of active star formation that have  been well surveyed in multiple wavelengths \citep{bachiller84,bachiller90,yan98,jor06,reb07,hiramatsu10}, and they are host to multiple known outflows including B1-a,b,c, and the outflows powered by IRAS~03304+3100, IRAS~03295+3050, IRAS~03292+3039, and IRAS~03282+3035 \citep{bachiller91,hirano97,degreg05,wal05B1,jor06,davis08}.

We identify three possible new outflows (or extension to previously known outflows)
 in the outer parts of the B1-Ridge (CPOCs 24, 26, 28) and eight located primarily around the periphery of the central B1 region (CPOCs 29 to 36) (see Figures~\ref{fig_reg3} and \ref{fig_reg4}). We mostly detect redshifted CPOCs in this region as there is a cloud along the same line of sight, at blueshifted velocities with respect to the B1 clump, that hinders our ability to detect outflow-related high-velocity blueshifted emission around B1.
Some of the CPOCs in this area (e.g., CPOCs 24, 26, 28) are probably extensions of outflows originating in the core of B1 and the B1-Ridge, just like the many HH objects recently found in the periphery of these star-forming regions \citep{wal05B1}. Other CPOCs found here could be new outflows associated with
young stars that lie outside the central core of B1 (e.g., CPOCs  31, 34, 35, and 36). With the relatively low resolution of our map it  
is difficult to concretely determine the driving source for  candidate outflows in this region of high protostellar density. We still give possible candidate 
sources (based on the proximity of the CPOC to a candidate YSO) in Table~\ref{cpoctab}.
Our discovery of new outflow candidates in the 
outer limits of these clusters show  
yet another example of how typical studies usually overlook the periphery of active star forming regions, and the need for studies that encompass a larger area in order to obtain a complete census of the outflows in the cloud. 



\subsection{Area V}
 The western part of Area V contains a scantly studied region of the Perseus molecular cloud complex 
 that lies between the B1 dark cloud and IC~348 cluster, and
  the eastern end includes the region surrounding the young stellar cluster IC~348 
 (see Figure~\ref{fig_reg5}). The region 
 east of B1 and west of IC 348  does not contain any major clusters or groups of young stars, but it does include 
 signposts of star formation. It  is host to three, poorly known, dark clouds (i.e., L1468, B3 and B4), several high-extinction cores \citep{wood94}, a
 few c2d YSO candidates \citep{evans09}, and IRAS sources \citep{lad93}.
  This area of Perseus is also the location of a warm shell of dust most likely the result of an \ion{H}{2} region associated with the early-B star HD~278942 that lies behind the cloud \citep{anderson00,rid06ring}.  

We identify eight new outflow candidates in this region (CPOCs 37 to 44), most of which are clustered about 1 degree to the west of IC 348.  CPOC 39 has a very elongated morphology, similar to the structure of a collimated outflow lobe, and it is located between the c2d YSO candidates SSTc2dJ033915.81+312430.7 and SSTc2dJ034001.49+311017.3. 
Although the latter YSO candidate is closer to CPOC 39, either of these two young stars  could be the powering source of this high-velocity gas.
We also find two sets of adjacent red and blue CPOCs that seem to pair to form two candidate bipolar flows. One of them is comprised of CPOC 40 and 42, and it is very close to IRAS~03363+3027 (see Figure~\ref{fig_reg5}). Although the IRAS source does not lie in the center of the two CPOCs, the uncertainty in its position (i.e., an ellipse with a major axis of 43\arcsec, minor axis of 13\arcsec, and a PA=73\arcdeg) is large enough that it is feasible that  IRAS~03363+3207 could be
the source of this candidate bipolar flow. The other, weaker, candidate  bipolar outflow is comprised of CPOC 37 and 38 (see Figure~\ref{fig_reg5}). We cannot assign a possible source to this candidate outflow as there are no known YSOs in this area. One possible reason for the lack of YSO candidates in the northwestern part of this region (where we find CPOCs 40 to 43, and  CPOCs 37 and 38) is the fact that this area was not covered by the c2d Spitzer IRAC observations \citep{jor06}.
\citet{wood94} identified a few cores just to the north of these candidate outflows and outside the northern edge of our CO map.
Hence, it may be possible that the fast moving gas observed in the northwest part of Area V (i.e., CPOCs 37, 38, 41, 43) is powered by undetected YSOs in the northern parts of Perseus.  We find CPOC 44 about 30\arcmin \/ southwest of the main IC 348 cluster, among a group of c2d YSO candidates 
(see Figure~\ref{fig_reg5}). 
Any of these young stars could be the powering source of this
high-velocity feature.

IC~348, at the eastern edge of Area V,  consists of two portions: the northeast part which contains the well known cluster of young stars; and the southwest part (about 10\arcmin \/ southwest of the center of the main cluster) which harbors a more recent burst of star formation activity \citep{herbst08}.  The young cluster located in the northeast portion of the region has been recently studied by a number of authors \citep{luhman03,muench07}, which show that the few hundred members of this cluster have a wide spread in age, ranging from Class I to Class III sources.  This portion of the region contains no known molecular outflows, and is home to a B5 star (HD 281159) that powers a  reflection nebula seen both in the optical and NIR.  
 The region to the southwest of the main cluster (also known as IC~348-SW; Tafalla et al.~2006)
contains high-density gas, has a higher extinction and harbors more embedded protostars than the rest of IC 348. It also contains a number of HH objects and a few well-known molecular outflows \citep{eisloffel03,tafalla06,walawender06}, including HH~211 \citep{mccaughrean94,gueth99}.  

We identify 4 potential new molecular outflow candidates located just to the east of IC~348-SW:
 CPOCs 45, 46, 48, 49 and 50  (see Figure~\ref{fig_reg5}). In the northern portion
of the main stellar cluster we also detect another candidate outflow, CPOC 46.  All of these CPOCs lie in a region rich in c2d YSO candidates, however our CO maps lack the necessary angular  resolution to confidently assign a powering source to each candidate outflow. 
One particular pair of high-velocity blobs is made of the red and blue CPOCs 48 and 49, which 
are located next to each other and might comprise the two lobes of a bipolar outflow. A candidate young star that could be the powering source of this candidate outflow is the nearby c2d YSO candidate  SSTc2dJ034458.55+315827.1.
In the IC~348-SW region we detect the high-velocity gas associated with HH~211 and other neighboring outflows, but we fail to detect any new outflows or outflow extensions.

\subsection{Area VI}
\label{reg6}

Area VI includes the  Barnard 5 (B5) dark cloud, also known as L1471, located at the eastern end of the Perseus molecular cloud complex.  
The region has only a few young stellar objects,  as revealed from infrared studies of the region 
 \citep[e.g.,][]{beichman84, evans09}. The c2d Spitzer IRAC survey did not cover the entire B5 region thus there may be a few low-luminosity YSO candidates in the eastern part of B5 not detected by IRAS. 
  The region is home to the well studied parsec-scale outflow from B5-IRS1 \citep{goldsmith86,fuller91,bal96b5,yu99}.

In this region we identify two high-velocity features which we suggest are new extensions 
of the B5-IRS1 outflow (CPOCs 51 and 59).
\citet{yu99} place the ends of each outflow lobe of the B5-IRS1 outflow at the position of V-shaped (i.e., bow-shaped) CO blobs  
at approximately $20\arcmin$ from the source, which they labeled C1 and C2. The redshifted CPOC 51  has an elongated structure, with a major axis that is coincident with the redshifted lobe of the B5-IRS1 outflow, and lies
 about $11\arcmin$ (or 0.8 pc at a distance of 250 pc) southwest of C2. The blueshifted CPOC 59 lies about $10\arcmin$ northeast of C1 and it is elongated in the direction towards B5-IRS1.  
The position, velocity and morphologies of the CPOCs 51 and 59 suggest they are part of the B5-IRS1 outflow and are associated with ejections that occur much earlier than those related to the C1 and C2 structures reported by \citet{yu99}. With these new extensions, the B5-IRS1 outflow has a total projected length
 of about $1\arcdeg$ (4.4~pc), approximately 50\% longer than previously thought. 
   This finding implies that there could be many other molecular outflows that are much larger than previously thought 
   (as shown for HH flows  by, e.g., Reipurth et al.~1997). It is possible that a significant number of the CPOCs reported here are extensions of
   previously known outflows, but in most cases it is hard to assign a source to the high-velocity blob (as discussed above, in \S~\ref{sourceid})

In this region we also detect nine other new candidate outflow features: 
CPOCs  47, 52, 53, 54, 55, 56, 57, 58, and 60 (see Figure~\ref{fig_reg6}). 
The most massive and energetic of these are CPOCs 55 and 56 (see Table~\ref{cpoctab}). We classify CPOC 56 as a candidate outflow as it:  has velocities that are far enough from the cloud to be considered a genuine high-velocity feature (see Figure~\ref{cpoc_vel}); 
  it is  within 5 to $10\arcmin$ of HH objects and a YSO;  
  and it shows a velocity structure where the gas at higher velocities is further away from the cloud center (similar to the so-called ``Hubble-law'' in molecular outflows where the outflow velocity increases with distance from the source). CPOC 55 lies very close to the YSO B5-IRS4 as well as several HH objects (see Figure~\ref{fig_reg6}) and shows a clearly different  structure from that of the cloud at redshifted velocities.   
 CPOC 47 is located just to the north of IC~348 where there are a number c2d YSO candidates, and this candidate outflow is most
 probably associated with one of these sources rather than any of the sources in B5. CPOC 52 is a blob with relatively high-velocity blueshifted gas, significantly different from ambient cloud velocities (see Figure~\ref{cpoc_vel}).
 CPOCs 53 and 54 have redshifted velocities  and may be associated with HH 844 and IRAS~03439+3233 (also known as B5-IRS3). CPOC 57 is redshifted and is located about 10\arcmin \/ northeast of B5-IRS4, while 
  CPOC 58 is located  south of the blueshifted lobe of B5-IRS1 and it is not clear to which young star in the region it is associated with. 
 CPOC 60 is located at the eastern edge of our map. We classify it as a candidate outflow because of its morphology and velocity structure.

\end{document}